\documentclass[a4paper,11pt]{article}
\pdfoutput=1 
\usepackage{jcappub}
\usepackage[T1]{fontenc}

\newcommand{\Msun}{$\rm M_{\odot}$}
\newcommand{\Mcrit}{${M_{\rm crit}}$}
\newcommand{\JLW}{${J_{\rm LW}(z)}$}
\newcommand{\Htwo}{$\rm {H_{2}}$}

\newcommand{\vbc}{$v_{\rm bc}$}
\newcommand{\sigvbc}{$\sigma_{\rm vbc}$}
\newcommand{\tdelay}{$t_{\rm delay}$}
\newcommand{\overdense}{$\delta(\vec{x})$}
\newcommand{\MIII}{$M_{\rm III,200}$}
\newcommand{\SMD}{$\rm SMD_{II}$}
\newcommand{\zcross}{$z_{\rm cross}$}
\newcommand{\error}{$\epsilon_{\rm NN}$}
\usepackage{breqn}
\usepackage{amsmath}
\usepackage{graphicx}
\usepackage{units}
\usepackage{sidecap}
\usepackage{float}
\sidecaptionvpos{figure}{c}
\usepackage{longtable}
\usepackage{multirow}
\usepackage{amsfonts}

\defcitealias{Paper1}{CF24}

\title{From Dark Matter Minihalos to Large-Scale Radiative Feedback: A Self-Consistent 3D Simulation of the First Stars and Galaxies using Neural Networks}

\author[a]{Colton R. Feathers}
\emailAdd{colton.feathers@rockets.utoledo.edu}
\author[a,b]{Mihir Kulkarni}
\emailAdd{mihir.kulkarni@uni-goettingen.de}
\author[a]{Eli Visbal}
\emailAdd{elijah.visbal@utoledo.edu}
\affiliation[a]{University of Toledo, Department of Physics and Astronomy and Ritter Astrophysical Research Center, 2801 W. Bancroft Street, Toledo, Ohio 43606}
\affiliation[b]{Institut für Astrophysik, Georg-August Universität, Friedrich-Hund-Platz 1, D-37077 Göttingen, Germany}

\abstract{A key obstacle to accurate models of the first stars and galaxies is the vast range of distance scales that must be considered. While star formation occurs on sub-parsec scales within dark matter (DM) minihalos, it is influenced by large-scale baryon-dark matter streaming velocities (\vbc) and Lyman-Werner (LW) radiative feedback which vary significantly on scales of $\sim$100 Mpc. We present a novel approach to this issue in which we utilize artificial neural networks (NNs) to emulate the Population III (PopIII) and Population II (PopII) star formation histories of many small-scale cells given by a more complex semi-analytic framework based on DM halo merger trees. Within each simulation cell, the NN takes a set of input parameters that depend on the surrounding large-scale environment, such as the cosmic overdensity, \overdense, and \vbc\ of the cell, then outputs the resulting star formation far more efficiently than is possible with the semi-analytic model. This rapid emulation allows us to self-consistently determine the LW background intensity on $\sim$100 Mpc scales, while simultaneously including the detailed merger histories (and corresponding star formation histories) of the low-mass minihalos that host the first stars. Comparing with the full semi-analytic framework utilizing DM halo merger trees, our NN emulators yield star formation histories with redshift-averaged errors of $\sim$7.3\% and $\sim$5.2\% for PopII and PopIII, respectively. When compared to a simpler sub-grid star formation prescription reliant on halo mass function integration, we find that the diversity of halo merger histories in our simulation leads to enhanced spatial fluctuations, an earlier transition from PopIII to PopII dominated star formation, and more scatter in star formation histories overall.}

\begin{document}
\maketitle
\flushbottom

\section{Introduction} \label{Intro}
The initial era of star and galaxy formation in the Universe is crucial to our understanding of cosmic evolution. Beginning $\sim$100 Myr after the Big Bang, the first stars are predicted to have formed within dark matter (DM) ``minihalos'' with masses below the atomic cooling limit of hydrogen (i.e. $M_{\rm vir} \simeq 10^5 - 10^6$ \Msun). The primordial gas clouds within these halos radiated energy via \Htwo\ molecular transitions, which limited gas cloud fragmentation as these transitions could not cool the gas as efficiently as metals. Simulations predict that this resulted in much larger characteristic stellar masses \citep[$M_{*} = 10 - 1000$ \Msun, e.g.,][]{Abel02, Bryan14-ENZO, Hirano14} than what is observed in the local universe \citep[see][for a recent review]{Klessen&Glover23}. These Population III (PopIII) stars lived relatively short lives of a few Myr \citep{Schaerer02} and critically initiated the processes of cosmic reionization and metal enrichment \citep[e.g.][]{Smith15, Chiaki13, Chiaki17}.

Since they form in small numbers and at high cosmic redshifts \citep{Skinner20}, PopIII stars have eluded direct observations thus far. While observations by \emph{JWST} have started hinting at possible PopIII sources \citep[e.g.][]{Naidu22, Labbe23, Finkelstein23}, many previous works have attempted to constrain the properties of first stars using indirect observations. For example, studies in ``stellar archaeology'' examine local extremely metal-poor stars to place lower limits on the PopIII initial mass function \citep[e.g.][]{Frebel15, Hartwig15, deBennassuti17, Graziani17, Griffen18, Magg18, Hartwig18b, Lorenzo22}. The optical depth of cosmic microwave background (CMB) radiation due to electron scattering has similarly been used to place upper limits on the efficiency of PopIII star formation \citep{Haiman03, Shull08, Ahn12, Visbal15b, Miranda17}. Further, line-intensity mapping of the 1640 \r{A} He II recombination line has been proposed as a method to constrain PopIII stellar properties and calibrate models of star formation \citep{Visbal15a, Parsons22-HeII}. Other studies have focused on the observability of PopIII pair-instability supernovae \citep{Whalen13, Hartwig18a} and PopIII gamma-ray bursts \citep{Bromm07, Burlon16, Kinugawa19} to constrain their properties. Lastly, the cosmological 21-cm signal is sensitive to the star formation rate density (SFRD) and three-dimensional (3D) clustering of the first stars and galaxies \citep[see][for a detailed review]{Pritchard12}. This is true for both global observations such as EDGES \citep{Bowman18-EDGES} and LEDA \citep{Price18-LEDA}, as well as those aiming to measure 3D spatial fluctuations such as HERA \citep{DeBoer17-HERA} and SKA \citep{Mellema13-SKA}. 

Theoretical predictions are necessary to guide and interpret these observations. One of the largest hurdles in accurately modelling the first stars is the vast range of distance scales that must be considered \citep[for detailed reviews on simulations of the first stars and galaxies, see][]{Greif15, Wise19}. While the process of star formation occurs on distance scales of parsecs (e.g.,  the virial radius of a $\sim 10^5\ \mathrm{M_{\odot}}$ halo at  $z=40$ is  $\sim 40$ pc), relative baryon-DM streaming velocities (\vbc) fluctuate significantly on 100 Mpc scales \citep{Tseliakhovich10-vbc, Tseliakhovich11} and can significantly hinder star formation as matter advects out of the shallow gravitational potential wells of the DM halos \citep{Greif11, Fialkov12, McQuinn12}. Radiative feedback from star formation can similarly affect cosmological volumes as ultraviolet (UV) radiation within the Lyman-Werner (LW) band (${E_{\rm LW}}$ = 11.2--13.6 eV) may freely stream up to $\sim$100 Mpc before cosmological expansion redshifts the photons into a Lyman series line and it is absorbed \citep{Haiman97,Ahn09}. This feedback may shine on to other DM halos, dissociating \Htwo \ molecules within them and further suppressing PopIII star formation \citep{Haiman97, Haiman00, Machacek01, O’Shea08, Ahn09}. To accurately model early star and galaxy formation then, one must simultaneously capture the star formation occurring on small spatial scales as well as the large-scale \vbc\ and radiative feedback.

In an effort to address this challenge, many works have analytically calculated the formation and large-scale distribution of the first stars to make predictions of observables and constrain their properties \citep[e.g.][]{Haiman06, Wyithe07, Dave12, Furlanetto17, Furlanetto22}. For example, several works have implemented the method of analytic halo mass function integration as a means of rapidly determining star formation in a given volume of space \citep[e.g.][]{Visbal15a, Visbal15b, Mashian16}. This method, however, requires fine-tuned parameters, such as the halo mass ranges over which one integrates to yield accurate SFRD evolutions when compared with more sophisticated calculations \citep[][henceforth \citetalias{Paper1}]{Paper1}. Further, this method can only predict statistically-averaged star formation for a given halo mass function at a given time, ignoring any previous star formation, radiation, gas heating and ejection, etc. that are required to accurately model the spatial fluctuations of star formation in non-global scenarios.

Semi-analytic models (SAMs) are another class of models that introduce more complex astrophysics on top of analytic frameworks to more accurately predict star and galaxy formation \citep[e.g.][]{Crosby13, Dayal20, Hegde23, Ventura24}. These typically implement DM halo merger trees to serve as the scaffolding for star and galaxy formation, and can include processes like self-consistent LW feedback \citep{O’Shea08, Agarwal12, Kulkarni13}, the inclusion of  relative streaming velocities \citep{McQuinn12}, and models for more complex halo physics induced by mergers, metal enrichment, and radiation \citep[e.g.][]{Magg18, Ahn21, Liu20}.

These SAMs, along with simpler analytic models, may be used within large 3D semi-numeric models to simulate the star formation within a subregion of the overall volume with the goal of reproducing observables that require spatial fluctuations and correlation \citep[e.g.][]{Visbal12, Visbal20, Munoz22, Munoz23, Cruz24}. Such semi-numeric models therefore more accurately predict the 3D distribution of high-redshift star formation and are typically used to both study the effects of the first stars on large-scale observables and constrain their properties \citep{, Fialkov13, Fialkov14a, Fialkov14b, Kaur2022, Reis22}. Note, semi-numeric models like the one presented in this work are distinguished from SAMs by including small-scale resolution elements in which an assumed sub-grid prescription dictates star formation. Such semi-numeric models, however, are often limited by computational resources and finite spatial resolution, meaning they typically cannot resolve the DM minihalos in which early star formation occurs while simultaneously capturing the large-scale evolution of the volume.

Recently, Magg et al. (2022) \cite{Magg22} elevated this sub-grid technique by implementing the semi-analytic framework A-SLOTH \citep{Hartwig22} to characterize the transition redshift between PopIII and PopII star formation in terms of cosmic overdensity, \overdense, then showed how the 21-cm signal changes with transition redshift. The authors utilized the A-SLOTH SAM to simulate star formation in N-body merger trees for a range of critical temperature thresholds and recovery times between the first PopIII supernova and the onset of PopII star formation. The results of these realizations were then used to populate cells of their 21-cm signal model with star formation based on the cell overdensities. While this represents perhaps the best effort to simultaneously account for both large and small distance scales in the literature to date, the authors did not self-consistently account for the effects of LW radiation and \vbc\ in their star formation model, only incorporating them into their 21-cm signal calculation. 

Here we present a new semi-numerical framework that self-consistently predicts the 3D distribution of PopIII and PopII star formation within a representative large-scale volume of the early universe. Our framework is the first to self-consistently compute large-scale effects, such as 3D spatial fluctuations in LW background intensity, while simultaneously capturing realistic DM halo merger trees that resolve the hosts of the first stars. We achieve this by developing many neural network emulators which reproduce the star formation results of a more computationally demanding SAM. Once trained, these neural networks rapidly and accurately emulate the star formation within a small-scale cell of our simulation volume while taking into account the larger scale environment of the cell (e.g., the specific LW intensity observed in that cell from the rest of the large-scale volume).

Machine learning techniques such as neural networks have recently been implemented in various fields of astrophysical research to efficiently explore large and complex datasets. For example, they have been utilized to emulate the distribution of galaxy halo masses and velocities \citep{Villanueva-Domingo22}, the 3D clustering of galaxies \citep{Bonici24}, the connection between galaxy formation and the DM halos which host them \citep{Behera24, Chittenden24}, overall structure formation \citep{Jamieson24}, gravitational lensing \citep{Hezaveh17}, and even the cosmic 21-cm signal to both constrain its parameters for future observations \citep{Kern17, Shimabukuro18} and identify the type of radio background present within it \citep{Sikder24}. Recognizing their potential, we implemented machine learning techniques in the semi-numeric framework presented here to rapidly populate sub-grid regions in our simulation with star formation. 

The rest of this paper is organized as follows. In Section \ref{Methods} we discuss the physical processes and parameters included in our simulation framework, and illustrate the neural net training process. In Section \ref{Results} we discuss the results of our model, and place them into context with other simulations. Finally, we summarize our results in Section \ref{Conclusions} and discuss planned future work with this new large-scale model. Unless otherwise stated, all distance scales are in comoving units and all baryon-DM streaming velocities are those found at Recombination. Also note that the \JLW\ values referred to in this work have units of $J_{\rm 21} = 10^{-21}\ \rm erg\ s^{-1}\ cm^{-2}\ Hz^{-1}\ sr^{-1}$. Finally, we assume a $\Lambda$CDM cosmology throughout, consistent with \cite{PlanckCollaboration}, adopting the following parameter values: $\Omega_{\rm m} = 0.32$, $\Omega_{\rm \Lambda} = 0.68$, $\Omega_{\rm b} = 0.049$, $h = 0.67$.

\section{Simulation Methods} \label{Methods}
We begin this section with a brief overview of our large-scale simulation and how it builds upon the work of \citetalias{Paper1}. We then describe the training procedure for our NN models to emulate the star formation within $\rm (3\ Mpc)^3$ resolution elements of the overall volume, taking into account their unique merger histories and environments. We conclude this section by explaining how we evolve the large-scale 3D volume through cosmic time in terms of star formation and feedback.

\subsection{Simulation Framework Overview} \label{Overview}
Within our semi-numeric framework, we implement artificial neural networks (NNs) via PyTorch to emulate both PopIII and PopII star formation in small-scale cells, given the halo merger history and environmental conditions of the cell (e.g., \vbc, \JLW, \overdense). We construct a simulation volume that is 192 Mpc on a side; this distance scale was chosen partially to save on computational resources allowing for multiple rapid realizations, and also because it encapsulates nearly twice the LW horizon distance allowing for more accurate \JLW\ calculations. We discretize the simulation volume into $64^{3}$ cells, each 3 Mpc on a side, which represents the distance scale over which the relative baryon-DM streaming velocity can be assumed to be constant \citep{Tseliakhovich10-vbc}. We then generate initial conditions for the cosmic overdensity, and \vbc\ of each cell using a perturbation-based approach, and populate each with a halo merger history corresponding to its \overdense. 

To set the star formation within our simulation cells, we implement a Monte Carlo merger tree-based SAM following the finite volume model of \citetalias{Paper1} (see their Section 6). In that model, we estimated the SFRD at each redshift step by cycling through all halos present at that time and introducing PopIII star formation into pristine halos with masses larger than \Mcrit(\JLW, \vbc)\footnote{Recently, \cite{Nebrin24} pointed out the potential importance of Lyman-$\alpha$ feedback in simulations of the first stars and galaxies. We note that this feedback is effectively included in our star formation efficiency parameters as the uncertainty in these values often outweigh the effects of such processes. In future work, we intend to find the most accurate values of the star formation efficiency as predicted by simulations which include all of the relevant physical effects.}. We calculated this critical mass using the model presented in \cite{Kulkarni21}, which was calibrated to hydrodynamical simulations including the effects of both LW and \vbc\ \citep[also see][for related \Mcrit\ models]{Schauer21, Nebrin23}. Following the onset of PopIII star formation, we imposed a 10 Myr recovery period before allowing halos to form PopII stars, as this simulated the time needed for PopIII star formation, feedback, and for the halo gas to re-cool. This ultimately resulted in PopII and PopIII SFRD histories for a $(3 \rm\ Mpc)^{3}$ volume at average cosmic density.

The training SAM used here is identical to the finite volume model of \citetalias{Paper1} except that we no longer include gas ejection and fallback due to supernovae in this work. We find that this removal results in $\sim 15\%$ drop in the final ($z = 15$) PopII SFRD, while having a negligible effect on the PopIII SFRD. This was done to allow for much more efficient training of NN models as the SAM no longer needed to track the gas ejected from each halo, nor the future descendant halo onto which the gas would fall back. By running each halo merger history through the $(3 \rm\ Mpc)^{3}$ SAM, we generate the expected star formation histories of those halos for a given combination of \JLW\ and \vbc\ (i.e. our ``training'' dataset). We then train our NN models on these results so that they may emulate the star formation within a simulation cell given the conditions and merger history of that cell. Then, once all emulators are trained, we begin stepping through time. At each time step, we determine the LW background intensity and \Mcrit\ of all cells, then cycle through each of them to emulate star formation where appropriate. 

\begin{figure*}
    \centering
    \includegraphics[width=\textwidth]{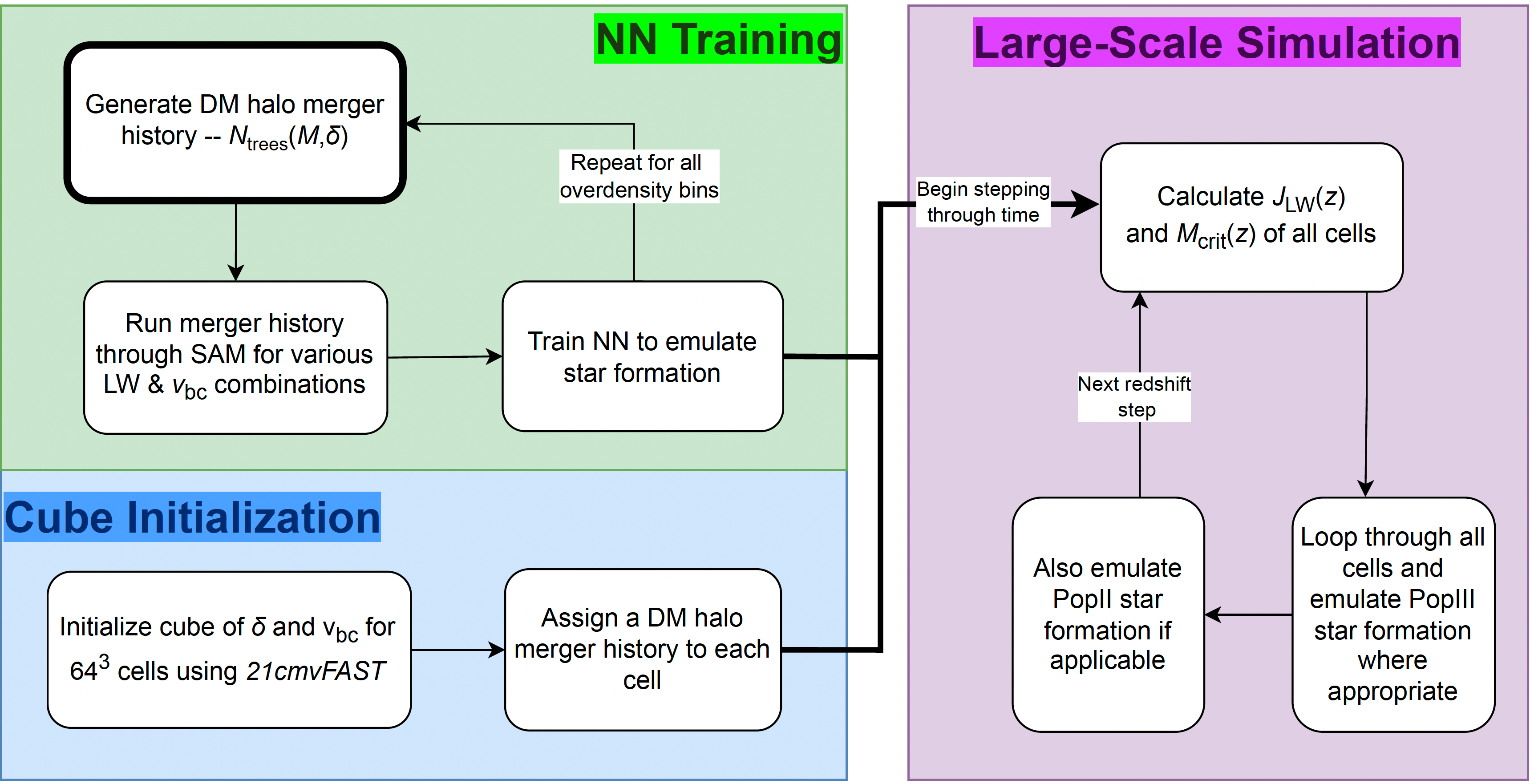}
    \caption{A simplified flowchart of our simulation framework. Starting with the thick bordered box in the green ``NN Training'' region, we generate a unique merger history for a given overdensity, $\delta$ (Section \ref{sub:Mergers}), and train it on the results of our SAM (Section \ref{sub:NN_Training}). We initialize the simulation volume of overdensities and \vbc\ using 21cmvFast, as described in \ref{21cmvFast} (blue region, ``Cube Initialization''). Finally, we begin stepping through time (purple region, ``Large-Scale Simulation''), determining the \JLW\ and \Mcrit\ of each cell (Section \ref{sub:LW-calculation}) at the current $z$. We then loop through each cell to emulate star formation where appropriate (Sections \ref{sub:PopIII-emulation} and \ref{sub:PopII-emulation}) before moving on to the next redshift step. See main text for further detail.}
    \label{fig:flow_chart}
\end{figure*}

The implementation of these NN emulators was necessary as the full SAM of \citetalias{Paper1} is too inefficient to serve as the star formation model of $64^{3}$ cells given the sheer number of DM halos we would need to simulate and cycle through. Thanks to their efficiency, our NN models can emulate the full star formation history of a given cell in an average time of roughly one second, whereas our SAM requires an average time of $\sim$200 seconds given the same environmental conditions. While the process of generating the training data sets and subsequently training the NN models can take up to $\sim$32,000 CPU-hours, our fully developed simulation can self-consistently simulate star formation between $z = 60 - 15$ for a $\rm(192\ Mpc)^{3}$ volume in $\sim$96 CPU-hours. The bulk of this runtime is devoted to the NN emulation, as $64^{3}$ NN models that each require one second to emulate star formation over $z = 60-15$ equates to $\sim$72 CPU hours. Most of the remainder of this runtime is dedicated to loading in the NN emulators and calculating the input parameters and \JLW. However, as this work represents a proof of principle that artificial NNs may be used to self-consistently link small-scale halo merger histories to large-scale radiative feedback, we mainly focused on finding an optimal set of training parameters to allow for accurate emulations of star formation and leave optimization of the NN architecture and training process to be the focus of future work (e.g. tuning of NN hyperparameters, more sparsely sampling the training parameter space, etc.). To aid the reader throughout the remainder of this section, we summarize our simulation framework in Figure \ref{fig:flow_chart} with a flow chart.

\subsection{Halo Merger Histories} \label{sub:Mergers}
We will now more carefully illustrate the procedures of our simulation framework, beginning with the DM halo merger history model. The merger trees used for producing the NN training data are generated following the prescription of \citetalias{Paper1}. We generate Monte Carlo merger trees based on Extended Press-Schechter formalism \citep{Press74, Bond91-EPSformalism}, assuming binary mergers of progenitor halos at each redshift step \citep{Lacey93}, separated by $\Delta z = 0.05$. Expanding upon \citetalias{Paper1}, we generate trees for 40 DM halo mass bins at $z = 15$, logarithmically spaced from $10^{5.6}-10^{9.5}$ \Msun, finding that this range encapsulates the vast majority of relevant halo masses for our simulation volume. Halos with masses $M_{\rm halo} > 10^{9.5}$ \Msun\ would only form in $\sim 0.5\%$ of all cells, while mass bins below $10^{5.6}$ \Msun\ do not exceed \Mcrit\ before $z=15$ and thus do not contribute to the SFRD at all. In total, we produced a library of over 95,000 DM halo merger trees which we use to assemble the unique merger histories that populate our simulation cells.

To determine the stochastic merger history for a particular overdensity (defined as \overdense\ $ \equiv (\rho(\vec{x}) - \bar{\rho}) / \bar{\rho}$ for all 3D positions, $\vec{x}$), we randomly determine the number of merger trees in each DM halo mass bin from a Poisson distribution about the $z = 15$ halo mass function. Note that, unless otherwise stated, all halo mass and overdensity values discussed are the values at $z = 15$. We calculate this halo mass function using the form presented in \cite{Barkana04} as it accounts for the effects that over/underdense regions have on the abundance of DM halos. This overdensity-dependent halo mass function is given by
\begin{equation} \label{EQ:HMF}
    \frac{dn}{dM}(M,\delta(\vec{x})) = \frac{\rho_{0}}{M} \left|\frac{dS}{dM}\right| (f_{\rm bias}),
\end{equation}
where $\rho_{0}$ is the cosmic matter density, $S(M)$ is the variance of the overdensity on regions with mass $M$, and the $f_{\rm bias}$ term encapsulates both the Sheth-Tormen and Press-Schechter halo multiplicity functions \citep[][respectively]{ShethTormen, Press74}. The Sheth-Tormen multiplicity function used here is the same as in \citetalias{Paper1}, and the extended Press-Schechter halo multiplicity function is adopted from \cite{Bond91-EPSformalism} such that 
\begin{align} \label{EQ:ST-EPS}
    \begin{split}
    f_{\rm bias}(\delta(\vec{x}),S) &= f_{\rm ST}(\delta_{\rm c},S) \times \left[\frac{f_{\rm PS}(\delta_{\rm c}-\delta(\vec{x}),S-S_{\rm R})}{f_{\rm PS}(\delta(\vec{x}),S)}\right] 
    \\ &= f_{\rm ST}(\delta_{\rm c},S) \times \biggl[ \frac{\delta_{\rm c}-\delta(\vec{x})}{\delta_{\rm c}} \left( \frac{S}{S-S_{\rm R}} \right)^{1.5} \mathrm{exp} \left(\frac{\delta_{\rm c}^{2}}{2S} - \frac{(\delta_{\rm c}-\delta(\vec{x}))^{2}}{2(S-S_{\rm R})}\right) \biggr] ,
    \end{split}
\end{align}
where $f_{\rm ST}$ and $f_{\rm PS}$ are the Sheth-Tormen and Press-Schechter halo multiplicity functions, respectively, $\delta_{\rm c} = 1.686/D(z)$ is the critical overdensity for halo collapse scaled to the present by the linear growth function, $D(z)$, and $S_{\rm R}(M)$ is the variance of an overdensity within a spherical volume equal to our simulation cell size extrapolated linearly to $z=0$. 

This hybrid approach for calculating the halo mass function is necessary because the Sheth-Tormen mass function used in \citetalias{Paper1} cannot be implemented for varying overdensities, unlike the Press-Schechter formalism with its conditional mass function. We therefore account for this by calculating the biased Press-Schechter halo mass function (bracketed term of equation \ref{EQ:ST-EPS}) and using it as a weighting function to adjust the Sheth-Tormen abundance values in our simulation \citep[see][for further detail]{Barkana04}. For each $\delta$ bin, we Poisson sample the number of trees for all 36 halo mass bins from the corresponding mass function, and assemble unique merger histories by randomly drawing that number of trees from our library of DM halo merger trees.

Note that, to further save on computational resources during the NN training process, we limit this Poisson sampling to 100 merger trees per halo mass bin. We find that the bin-averaged star formation rate (SFR) does not meaningfully change by adding more DM halo merger trees. In the simulation, the average SFR($z$) given by one of these limited mass bins is multiplied by the ratio of 100 to the Poisson sampled value to account for this truncation in merger trees. 

\subsection{Neural Net Model Training} \label{sub:NN_Training}
Due to the computational constraints discussed above, we require efficient and rapid determination of the SFR within a cell given its environmental conditions. To do this, we emulate star formation in each cell of our simulation volume using PyTorch to implement NN regression machine learning algorithms \citep[for a review on the use of machine learning techniques in cosmological simulations, see][]{Moriwaki23}. These models rapidly extrapolate sparsely populated, N-dimensional parameter spaces, allowing us to emulate star formation within the DM halos of a cell, given a set of relevant input parameters such as its \JLW\ and \vbc. 

Once initial conditions of the simulation volume are generated (detailed in Section \ref{21cmvFast}), we sort the cell overdensities into 200 evenly spaced bins between the minimum and maximum $\delta$ values of the entire box, then fit the distribution to a Gaussian. We descretize this Gaussian into 400 overdensity bins such that the area under the curve of each bin is equivalent. This binning procedure results in many overdensity bins near the peak of the distribution ($\Delta\delta_{\rm min} = 1.24 \times 10^{-3}$), and very few bins along the tails ($\Delta\delta_{\rm max} = 0.352$). 

For each overdensity bin, we generate a unique DM halo merger history from our tree library as discussed above, giving us a total of 400 unique merger histories that are used throughout the volume. Each simulation cell is then assigned with the merger history of the closest overdensity bin to its individual overdensity, representing the DM scaffolding for star formation within it. We note that populating $64^{3}$ cells with 400 merger histories likely results in neighboring cells sharing identical merger histories. However, we find that this is a rarity in our simulation, as only $\sim 2.58\%$ of all cells neighbor another cell with the same merger history. Despite this, we find that all quantities of interest are well converged with respect to the number of independent merger histories used in our simulation framework (discussed further in Section \ref{Results}).

For both PopII and PopIII star formation, we implement a four-layer NN model. The architecture of each model is set up to take in a predetermined number of inputs based on the stellar population being emulated, pass those through two hidden, fully-connected neural layers with 50 nodes each, then output the resulting star formation value for a given simulation cell at a given redshift. We implement the \emph{ReLU} activation function in training our NN models due to its computational efficiency for large training data sets, and optimize our model parameters with the \emph{Adam} algorithm. We train the PopII and PopIII star formation emulators of a given overdensity bin over $10^{3}$ and $10^{4}$ epochs, respectively. 

We note that this particular framework was chosen because of its relative simplicity and ability to rapidly and accurately emulate star formation. While we did not exhaust all possible avenues of NN architecture in our preliminary research, and other setups may further benefit the emulation accuracy, our chosen framework is able to reliably predict star formation for the purposes of this work.

To generate the input training parameters for our NN models, we run all 400 $\delta$-dependent merger histories through the $(3\ \rm Mpc)^{3}$ box SAM of \citetalias{Paper1} using many combinations of \JLW\ and the root-mean-square \vbc\ at Recombination, \sigvbc\ (which corresponds to $\rm \sim 30\ km\ s^{-1}$ at $z\approx 1060$, \cite{Tseliakhovich10-vbc}). We generate 86 realistic training LW background histories based on previous iterations of our final simulation, and use 13 bins of \sigvbc\ evenly spaced from zero to three, finding that this sufficiently samples the typical parameter spaces of both. We run all 400 DM halo merger histories through every combination of \JLW\ and \sigvbc, yielding 1118 realizations for each merger history on which the NN models train.

\subsubsection{PopIII Star Formation} \label{sub:PopIII-emulation}
We will now discuss the specific NN inputs needed to output PopIII star formation in our model framework. We define the value emulated for PopIII star formation as \MIII$(z)$, which represents the cumulative integer number of 200 \Msun\ PopIII star formation events that occur within a cell, i.e. \MIII$(z)$\ $\equiv (M_{\rm *,III}(z) / 200\ M_{\rm \odot})$ where $M_{\rm *,III}$ is the total PopIII stellar mass ever formed by redshift $z$. As in \citetalias{Paper1}, we assume that all PopIII star formation events result in 200 \Msun\ of stellar mass for simplicity. While more sophisticated treatments of primordial star formation may be implemented, such as randomly sampling from an initial mass function, this is outside the scope of this work and so we leave this for future iterations of our simulation framework.

Running all LW-\vbc\ combinations through the full SAM yields the raw \MIII$(z)$ histories which the PopIII NN emulators are then trained to output. At high redshifts, however, \MIII\ rises in discrete intervals as individual halos overcome \Mcrit, resulting in a piecewise-like behavior that the NN models struggle to reproduce. We have therefore found that logrithmically interpolating \MIII$(z)$ between subsequent values boosts the accuracy of our NN emulations as it relaxes the initially bursty growth into a more steady, ever increasing trend. This effectively trains the NN models to capture a more time-averaged number of DM halos overcoming \Mcrit, rather than the underlying DM halo merger history responsible for discrete bursts. However, this does mean that the \MIII\ value output by our NN models is also smoothed, and so we round down to the nearest integer when obtaining the PopIII SFR to account for this. This rounding is not performed, however, if the NN gives \MIII\ values between zero and one when first emulating, as this would give zero PopIII SFR and effectively delay star formation in the cell, leading to less accurate emulations for the remaining simulation.

To self-consistently emulate \MIII, our NN models take input parameters that depend on both present and past conditions of the host cell. For PopIII star formation, the NN models take five input parameters which we now discuss in detail: \textbf{(1)} current baryon-DM streaming velocity of the cell, \vbc$(z)$, which has units of km $\rm s^{-1}$ and, once initialized by 21cmvFast, gives our NN model key information to accurately emulate the critical mass for PopIII star formation. \textbf{(2)} The logarithm of the current LW background intensity, $\log_{10}$(\JLW) which, alongside the \vbc\ input, allows the NN models to accurately track the evolution of \Mcrit\ in each cell. \textbf{(3)} The logarithm of the current most massive halo (MMH) within the cell in units of \Msun, i.e. $\log_{10}$(MMH$(z)$). This input is crucial for accurately emulating the onset timing of PopIII star formation in each cell, since it will be the first halo to overcome \Mcrit\ at any given redshift step. \textbf{(4-5)} We integrate both \JLW\ and \Mcrit$(z)$ from the start of the simulation at $z = 60$ up to the most recent time step, $z_{\rm prev}$, and take the logarithm of each, i.e. $\log_{10}(\int_{60}^{z_{\rm prev}} J_{\rm LW}(z^{\prime}) dz^{\prime}$) and $\log_{10}(\int_{60}^{z_{\rm prev}}$ \Mcrit$(z^{\prime})\ dz^{\prime}$). These last two input parameters give the NN model information regarding the magnitude of previous PopIII star formation and radiative feedback experienced by the cell, quantities which constrain present star formation (e.g. cannot make more stars than available baryonic mass). Summaries of each NN model input parameter are listed in Table \ref{tableOparams}.

We note that, for both PopII and PopIII emulations, we take the logarithm of most input values to limit the dynamic range of the NN input parameters. Further, a zero intensity LW background would yield infinities when taking the logarithm of \JLW\ and its integral, and so we impose a ``primordial LW background'' everywhere in the simulation of $J_{\rm LW,0} = 10^{-6}\ J_{21}$ to avoid this. This extremely low background intensity serves as a LW floor below which no cell may fall, and allows efficient NN model training while not affecting \Mcrit\ throughout the box.

\subsubsection{PopII Star Formation} \label{sub:PopII-emulation}
We train our NN models to emulate the PopII stellar mass density (\SMD) using only two input parameters: \textbf{(1)} the logarithm of the cell \Mcrit\ value one delay time prior to the present, $\log_{10}(M_{\rm crit}(t(z)-t_{\rm delay}))$. \textbf{(2)} The logarithm of the integral of \MIII$(z)$ up to a delay time ago with respect to Hubble time, i.e. $\log_{10}(\int_{t(z=60)}^{t-t_{\rm delay}}$ \MIII$(z(t^{\prime}))\ dt^{\prime}$). We assume the delay period to be the same as in \citetalias{Paper1}, \tdelay\ = 10 Myr, representing the time needed for PopIII stars to exhaust their fuel, disrupt and enrich the halo gas with their stellar demise, and for the gas to cool once more and form PopII stars \citep{Jeon14}. The first input informs the NN model how many DM halos are above \Mcrit\ at that time, giving it an estimate for the number of halos that have already self-polluted with PopIII star formation by \tdelay\ ago, and would now be forming PopII stars. Similarly, the second input grants our NN models insight as to how much PopIII star formation had already happened up to one delay time ago. Since halos only experience PopII star formation after they undergo PopIII star formation and the subsequent time delay, and because each NN model is trained to replicate the star formation of a given merger history, these two inputs are sufficient to accurately emulate PopII star formation (see Table \ref{tableOparams} for a summary of input parameters for both PopII and PopIII NN models). 

\begin{table}[t]
    \centering
    \begin{tabular}{c c c}
        \hline
        Population & Output & Input Parameters\\
        \hline\hline
        
        \multirow{5}{2em}{PopIII} & \multirow{5}{2em}{\MIII} & \vbc$(z)$\\
        & & $\log_{10}$(MMH$(z)$)\\
        & & $\log_{10}$(\JLW)\\
        & & $\log_{10}(\int_{z=60}^{z^{\prime}}$ \JLW\ $dz$)\\
        & & $\log_{10}(\int_{z=60}^{z^{\prime}}$ \Mcrit$(z)\ dz$)\\
        \hline
        \multirow{2}{2em}{PopII} & \multirow{2}{2em}{$\rm SMD_{II}$} & $\log_{10}(M_{\rm crit}(t-t_{\rm delay}))$\\ 
        & & $\log_{10}(\int_{t(z=60)}^{t-t_{\rm delay}}$ \MIII$(z)\ dt$)\\         
        \hline
    \end{tabular}
    \caption{Training parameters for our NN models. From left to right, the columns denote the stellar population being emulated, the NN output parameter for each, and the various input parameters. We train our PopIII and PopII NN models over $10^{4}$ and $10^{3}$ epochs, respectively. See Sections \ref{sub:PopIII-emulation} and \ref{sub:PopII-emulation} for further details on specific inputs.}
    \label{tableOparams}
\end{table}

\begin{figure}
    \centering
    \includegraphics[width=\textwidth]{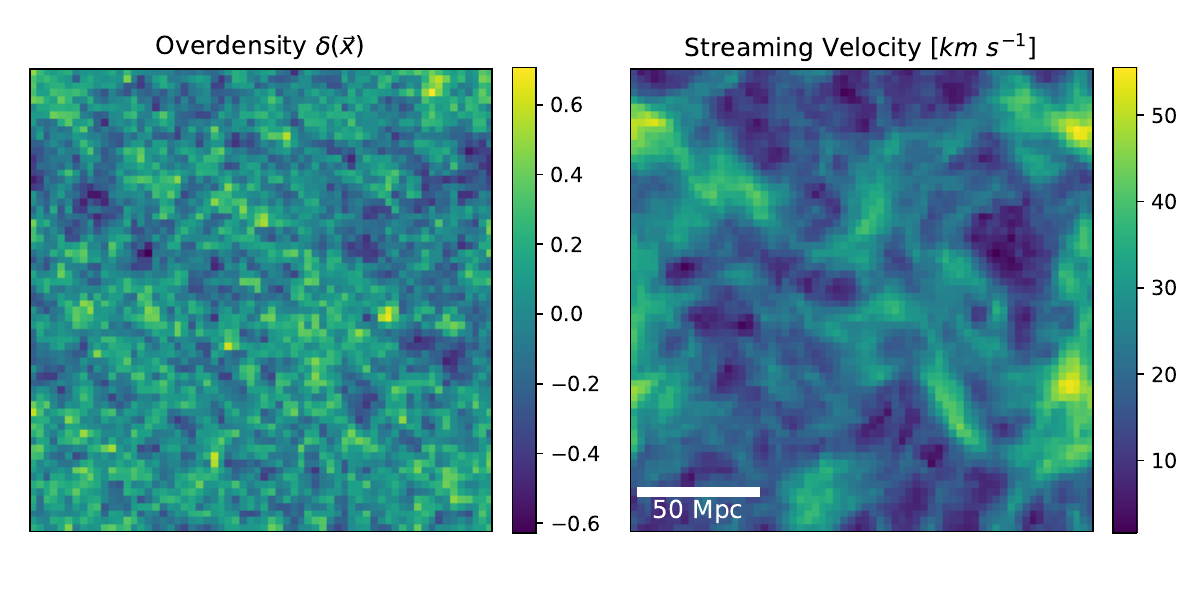}
    \caption{Central slices of our simulation volume, generated by 21cmvFast \citep{Munoz19}. \textbf{Left:} The cosmic overdensity value of each cell. \textbf{Right:} The relative baryon-dark matter streaming velocity of each cell. Note that the overdensity values are those at $z$ = 15, while streaming velocities are plotted in terms of km $s^{-1}$ at $z$ = 1060.}
    \label{fig:field-slices}
\end{figure}

\subsection{Full 3-Dimensional Simulation} \label{Full_Sim}
In this subsection, we discuss initialization of the full 3D simulation volume and summarize one full time step of the overall simulation. Once the PopII and PopIII NN emulators have been trained for all bins of $\delta$, our large-scale simulation can be summarized in five steps (also see Figure \ref{fig:flow_chart}): 
\begin{enumerate}
    \item Initialize conditions of the simulation volume, assigning each cell a \vbc\ and \overdense\ (along with the corresponding merger history). Then begin stepping through time starting at $z=60$
    \item  Calculate current LW background intensity and \Mcrit(\JLW, \vbc) for all cells
    \item Cycle through all $64^{3}$ cells to check if the MMH in each has overcome \Mcrit, emulating PopIII star formation if so
    \item If it has been $\geq$\tdelay\ since the onset of PopIII star formation, emulate PopII star formation in those cells as well
    \item Once finished emulating in all appropriate cells, advance to the next redshift step and repeat steps 2-5 until $z=15$
\end{enumerate}

\subsubsection{Initial Conditions} \label{21cmvFast}
The overdensities and baryon-DM streaming velocities of each cell in our simulation are initialized using 21cmvFAST \citep{Munoz19}. This cosmological framework calculates the 21-cm power spectrum with consideration for LW feedback and the 3D fluctuations in the matter density, but expands on the original 21cmFast \citep{Mesinger11-21cmFAST} to include the effects of \vbc\  \citep{Tseliakhovich10-vbc, Tseliakhovich11}. For our purposes, we utilize 21cmvFast to generate a $(192\ \rm Mpc)^{3}$ cube of $64^{3}$ cells with individual \overdense\ and \vbc\ values, and we show slices of these cubes used in our model in Figure \ref{fig:field-slices}. Note that the \vbc\ values shown are those at $z = 1060$, and the \overdense\ values are linearly extrapolated to $z = 15$, finding that linear growth sufficiently approximates the growth of fluctuations at our chosen redshifts. Once all NN model training is complete and we have generated the conditions of all cells in the simulation volume, each cell is assigned the halo merger history that corresponds to its \overdense\ and the simulation may begin.

\subsubsection{Lyman-Werner Background} \label{sub:LW-calculation}
At each time step in our simulation, we must determine the LW background intensity seen by each cell, \JLW. The intensity of this background depends on the degree of star formation occurring at a given lookback time, as radiation from this process streams across the volume and dissociates the \Htwo\ necessary to cool the halo gas for PopIII star formation. We assume that star formation occurs over one time step which translates to a spike in SFR over that step. To both avoid unphysical instantaneous bursts of LW and accurately model stellar lifetimes, we assume that the SFR of a given time step is evenly spread out over a time span of 3 Myr. We assume this value to be the lifetime of the star, consistent with the PopIII stellar lifetimes of \cite{Klessen&Glover23} for masses $\gtrsim$100 \Msun\ (see their Figure 8). During this time period, stellar radiation streams out of the cell across the simulation volume, contributing to the LW background. 

Note, to avoid the growth of unrealistic numerical feedback and also improve the stability of our NN emulators, we calculate the PopIII SFR of each cell by averaging its \MIII$(z)$ over ten simulated time steps ($dz =$ 0.5) to get $\bar M_{\rm III,200}(z)$. We then round this average value down to the nearest integer to account for the \MIII\ smoothing used in NN training (Section \ref{sub:PopIII-emulation}), then calculate the PopIII SFR of the cell as $SFR_{\rm III}(z) = (d \bar M_{\rm III,200} / dt) \times 200$ \Msun, where $dt$ is the elapsed time in years between $z$ and $z + \Delta z$. While the SFR smoothing discussed in the previous paragraph is implemented to mimic the lifetime of early high-mass stars, this \MIII\ averaging prevents large bursts of emulated star formation from boosting \JLW\ and \Mcrit, thereby suppressing star formation the following time step and initiating an unphysical oscillatory behavior in the PopIII SFR. If we do not average \MIII$(z)$ before calculating the PopIII SFR, these oscillations progressively affect the entire volume until all cells synchronously oscillate between time steps of zero star formation and LW feedback and steps of large bursts of both. We find such behavior to be unphysical, likely an effect of of the discretization of space and time when determining LW feedback, and is something we intend to research further and mitigate in future work. 

To calculate \JLW\ of every cell simultaneously, we smooth the PopII and PopIII SFRs over 20 predetermined comoving distance scales up to the LW horizon. This smoothing is achieved through the use of fast Fourier transformations of the raw SFR histories convolved with a spherical top hat window function over each distance scale. For a given cell, these concentric spheres represent the total SFR at different times in the simulation history, the radiation of each sphere only just now arriving at the central cell from their respective distances. We take the difference of the SFRs in subsequent spheres then divide by the difference in their volumes to get shells of SFRD  seen by each cell over time. From these 20 shells, we interpolate the SFRD at each simulation time step out to the LW horizon, and determine its intensity following the \JLW\ calculation of \cite{Visbal20}:
\begin{equation}
    J_{\rm LW}(z) = \frac{c(1+z)^{3}}{4\pi} \int_{z}^{\infty} \epsilon_{\rm LW}({z^{\prime}}) \Bigg|\frac{dt_{\rm H}}{dz^{\prime}}\Bigg| f_{\rm LW}(z^{\prime}, z)\ dz^{\prime}. 
\end{equation}
Here, $\epsilon_{\rm LW}({z^{\prime}})$ is the mean LW emissivity, $t_{\rm H}$ is the Hubble time, and $f_{\rm LW}(z^{\prime}, z)$ is the attenuation of the LW flux from $\rm {z^{\prime}}$ to $z$ as LW photons redshift into Lyman series lines and are absorbed \citep{Haiman97}. As in \citetalias{Paper1}, we approximate this attenuation using equation 22 of \cite{Ahn09}, and determine $\epsilon_{\rm LW}({z^{\prime}})$ by summing the contributions of both stellar populations.

\begin{SCfigure}
    \centering
    \includegraphics[width=0.55\textwidth]{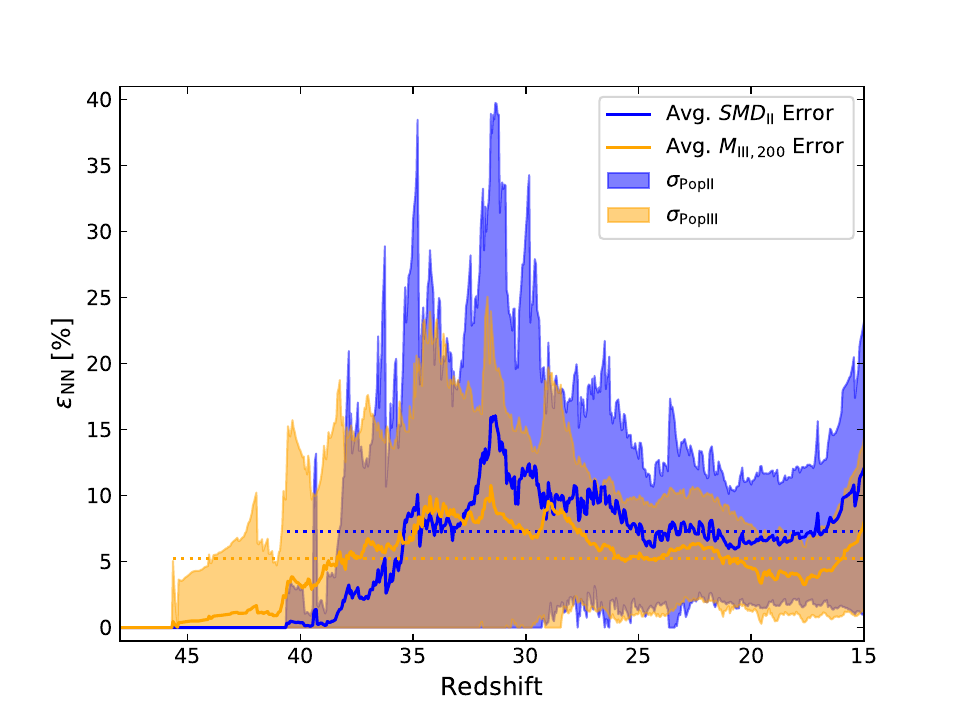}
    \caption{Average error of the emulated star formation values with respect to those of the full SAM for both PopII \SMD\ (blue) and PopIII \MIII\ (orange). We randomly sample 100 cells from our simulation and run their merger histories through the finite volume SAM of \citetalias{Paper1} using the simulated \JLW\ history, \overdense, and \vbc\ of each cell as the input parameters. We calculate \error\ at each redshift step via equation \ref{EQ:perc-diff}, and average these errors across all 100 sample cells to get the mean \error\ at each $z$ (solid lines) as well as the standard deviation (shaded regions). Finally, averaging \error$(z)$ across all redshifts with nonzero star formation gives averages of $\sim$7.3\% and $\sim$5.2\% for PopII and PopIII, respectively. These are represented by the correspondingly colored dotted lines which extend across the redshift range over which they are averaged.}
    \label{fig:Perc_Diffs}
\end{SCfigure}

We determine the \JLW\ contributions from all lookback times up to the LW horizon by performing this window function smoothing and interpolating the SFRD value at each intermediate step. We do this not only to negate the change in distance between subsequent redshift steps as the universe expands, but also to boost computational efficiency as the LW horizon encapsulates dozens of previous redshift steps from earlier in the simulation, and so we avoid smoothing over all of those individual distance scales. We find that this interpolation procedure reproduces the SFRD shells of smoothing over all previous redshift steps to the percent level while significantly reducing overall runtime. Once armed with the SFRD histories of each cell, we calculate the resulting \JLW\ and determine \Mcrit(\JLW,\vbc) for each cell using the model presented in \cite{Kulkarni21}.


\section{Results} \label{Results}
We now discuss the results of our simulation framework, first by illustrating the accuracy of our NN emulators. Next, we discuss the star formation results at the scale of the entire $\rm (192\ Mpc)^{3}$ volume and compare the averaged results to those of the global model presented in \citetalias{Paper1} as well as with a simpler modelling prescription reliant on analytic halo mass function integration. We then show how star formation changes with cell environment and simulation framework, before ending this section with a comparison of the spatial clustering of star formation between models.

\begin{figure}
    \centering
    \includegraphics[width=\textwidth]{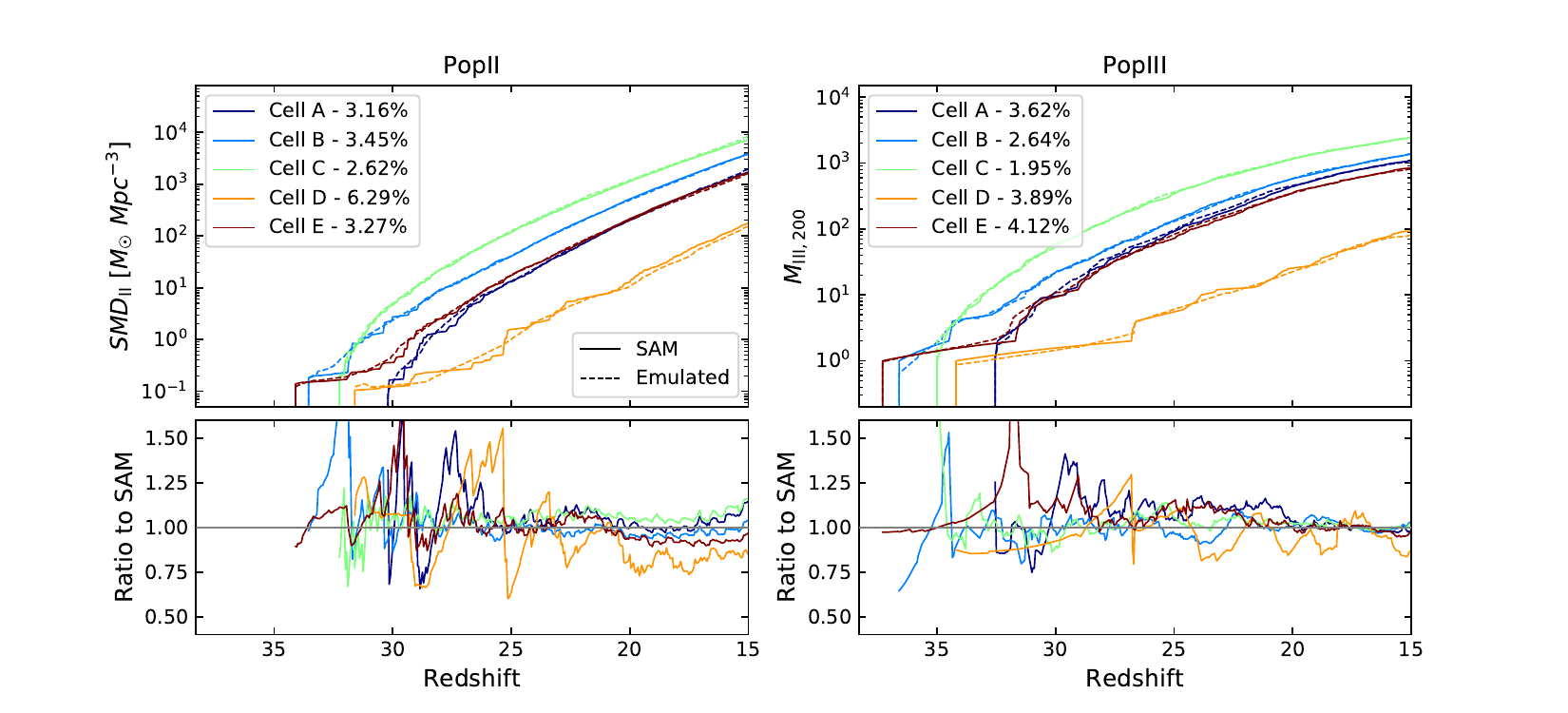}
    \caption{Sample PopII \SMD\ (left) and PopIII \MIII\ (right) histories of five individual sample cells from our simulation compared with the expected histories from our SAM framework. \textbf{Top Panels:} We compare the emulated star formation histories of our simulation (dashed) to those resulting from the finite volume SAM of \citetalias{Paper1} (solid), given the same cell conditions as the sample cell (see Section \ref{sub:Fid_Results} for further detail). Individual colors correspond to one sample cell, and we denote these as Cells A-E in the legend, followed by their redshift-averaged \error\ with respect to the SAM results (equation \ref{EQ:perc-diff}). \textbf{Bottom Panels:} The ratio of the emulated star formation histories to their respective SAM history in the top panels.}
    \label{fig:SAM_Emul_Int}
\end{figure}

\subsection{Neural Network Accuracy} \label{sub:Fid_Results}
To assess the accuracy of our NN emulations, we randomly sample 100 cells from our simulation and record their \JLW\ histories, cosmic overdensities, and streaming velocities. We then run each of these sample cells through the original semi-analytic framework using their corresponding halo merger history and environmental conditions to generate the expected star formation history of each cell (e.g. the expected \SMD$(z)$ and \MIII$(z)$). This effectively forms a ``test'' dataset for the NN models which we can use to calculate the error with respect to the emulated star formation history (i.e. the NN generated \SMD$(z)$ and \MIII$(z)$)\footnote{Note, we do not explicitly generate a ``validation'' dataset for tuning our NN emulators as our main purpose is to demonstrate the viability of such emulators to unite small-scale merger histories with large-scale radiative feedback. As with the NN input parameters, we tested various hyperparameter values through trial-and-error until the resulting emulation was sufficiently accurate. Our final training model uses a learning rate of $8 \times  10^{-5}$, a batch size of 100, and a weight decay of $10^{-5}$. Finer tuning of hyperparameters is saved for future work.}. We calculate the fractional error of a sample cell, \error, via 
\begin{equation} \label{EQ:perc-diff}
    \epsilon_{\rm NN}(z) = \frac{|SF_{\rm i,emulated}(z) - SF_{\rm i,SAM}(z)|}{SF_{\rm i,SAM}(z)},    
\end{equation}
where $SF_{i}$ represents either \SMD\ or \MIII.

We determine this error at each redshift step to get \error$(z)$, then average all 100 \error$(z)$ histories to get the mean fractional error of our NN emulation models over time. We plot these average errors as percentages along with their standard deviations in Figure \ref{fig:Perc_Diffs}. Note that we will refer to these errors in terms of percent error for the remainder of this work. From Figure \ref{fig:Perc_Diffs} we see that all randomly sampled cells yield emulated star formation values that are, on average, $\lesssim$15\%\ different from those of the full SAM, regardless of redshift or stellar population. We note, however, that the NN emulations often struggle to reproduce the initially bursty star formation of the merger trees, causing spikes in error. Further, as more cells form stars and their observed LW backgrounds diversify, the average percent error rises as the training LW backgrounds are less densely sampled at intermediate redshifts ($35 \lesssim z \lesssim 25$), and some cells with rarer conditions may reach percent errors of $\gtrsim$30\%. However, even this error is relatively small when compared to the $\sim$order of magnitude uncertainties in astrophysical parameters such as the star formation efficiency. Regardless, averaging these curves over all redshifts with star formation gives mean percent differences of $\sim$7.3\% and $\sim$5.2\% for PopII and PopIII, respectively, further emphasizing the overall accuracy of our NN emulators.

\begin{figure*}
    \centering
    \includegraphics[width=0.9\textwidth]{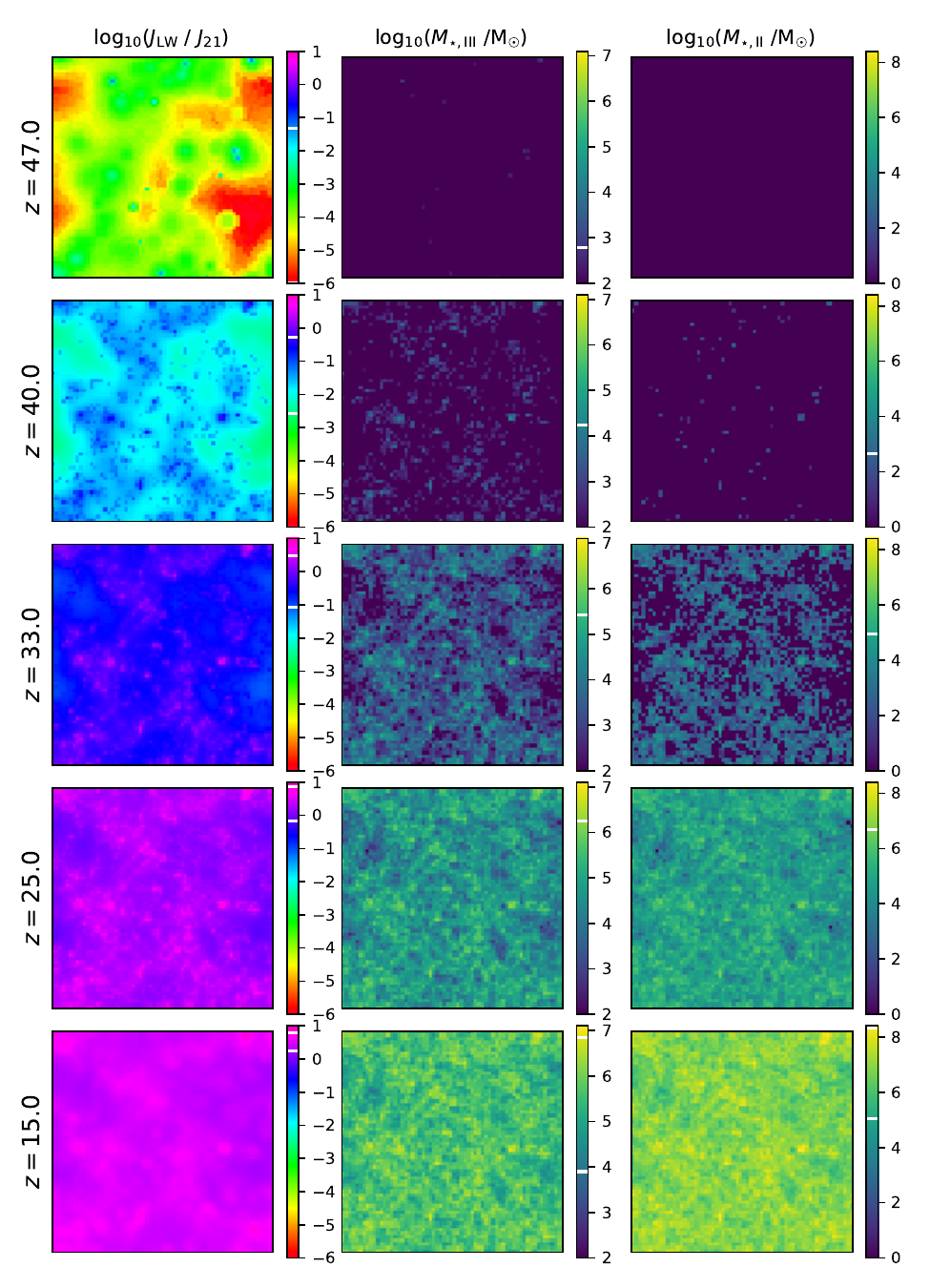}
    \caption{Slices of the LW background intensity (left column), along with the cumulative PopIII (center column) and PopII (right column) stellar masses at various redshift steps within our full semi-numeric simulation. Individual rows are labelled by redshift, and color bars for each panel are shown in logarithmic space. Within each color bar, the white lines denote the maximum and minimum value of the slice, showing the dynamic range of each.}
    \label{fig:Video_Shots}
\end{figure*}

To visualize this accuracy, we compare five emulated star formation histories from cells of our simulation to those resulting from our original SAM in Figure \ref{fig:SAM_Emul_Int}. From the right column, we see that the NN emulations very closely trace the full SAM PopIII star formation histories. The redshift-averaged \error\ of these five emulated \MIII\ curves range from $\sim$2.0$-$4.1\%, as noted in the legend. In the bottom panel, we see that the ratios of each emulation with respect to its corresponding SAM are centered around unity at most redshifts, and that these generally get more accurate with time. As in Figure \ref{fig:Perc_Diffs}, this is mainly a consequence of the initially bursty behavior of PopIII star formation. As a few individual halos begin overcoming \Mcrit, \MIII\ rises in discrete jumps which we smoothed in NN training to boost emulation accuracy (see Section \ref{sub:PopIII-emulation}). With time, \MIII\ begins to rise more consistently as halos begin to overcome \Mcrit\ at every time step, giving a star formation evolution that is much easier to accurately emulate. In the left panels of Figure \ref{fig:SAM_Emul_Int}, we again see tight agreement between the NN emulated and SAM PopII star formation histories. These emulations have average percent differences of $\sim$3.2$-$6.3\%, which is reflected in the bottom left panel as their ratios hover around unity for most of the simulation as well.

\subsection{Full Simulation Results} \label{sub:Comparison}
We now shift our focus to the overall results of our $(192\ \rm Mpc)^{3}$ simulation. We begin this subsection by presenting the star formation and of the LW background intensity evolution of the volume, then discuss more detailed comparisons between the volume-averaged trends of our simulation and the global SAM of \citetalias{Paper1}. We also compare these results to those utilizing a simpler analytic framework in place of our sub-grid NN emulation to show the impact that DM halo merger trees have on star formation.

\subsubsection{Evolution of Simulation Volume}
To visualize the evolution of the simulation volume, we plot slices of \JLW\ alongside the cumulative PopIII and PopII stellar masses at various redshifts in Figure \ref{fig:Video_Shots}. From top to bottom, we see the buildup of LW radiation and stellar mass throughout the volume as time progresses. Qualitatively, the diversity of DM halo merger histories in our simulation cells causes visibly enhanced small-scale fluctuations in the stellar mass when compared to the underlying density field in Figure \ref{fig:field-slices}, a feature we explore further in Section \ref{subsub:SFRs_Integral}. On larger scales, regions with low \vbc\ also tend to have higher star formation and \JLW\ due to having lower critical masses overall, and this is most pronounced in the LW column of Figure \ref{fig:Video_Shots}.

Almost every cell has begun star formation by $z=25$, and while it still traces the underlying density field, the dynamic range of stellar masses in Figure \ref{fig:Video_Shots} has decreased. Moreover, the \JLW\ dynamic range steadily falls from almost five dex at $z = 47$ to a factor of $\sim$3.6 at the end of the simulation. By $z =$ 15, not only has PopII star formation outpaced that of the PopIII, its smooth growth has further washed out large-scale differences in all three panels. The final PopII and PopIII stellar masses have similar dynamic ranges, being just over and under three orders of magnitude, respectively. 

\subsubsection{Comparison with Global Results} \label{subsub:Global_Compare} 
\begin{figure*}
    \centering
    \includegraphics[width=\textwidth]{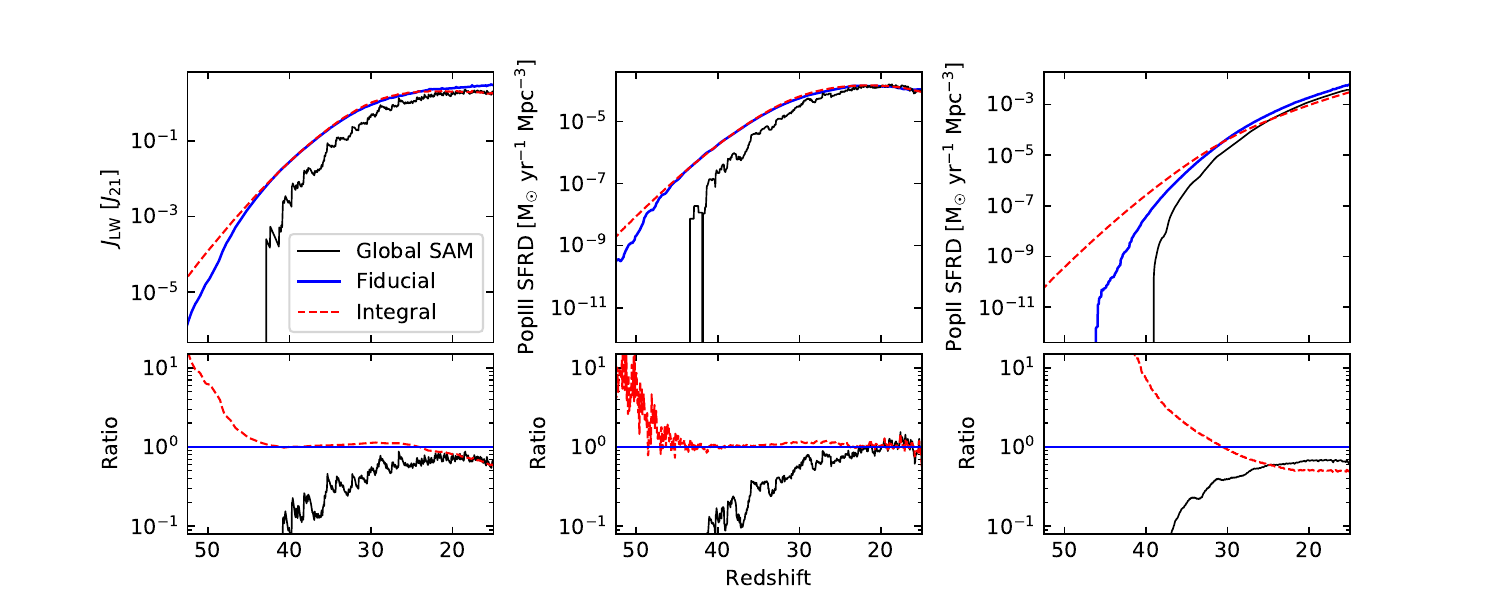}
    \caption{A comparison of the volume-averaged results of our full, NN-based simulation (solid blue) with that of the Integral method (dashed red) and the globally-averaged SAM of \citetalias{Paper1} (solid black). The top panels depict the average redshift evolution of \JLW\ (left), and both the PopIII (middle) and PopII (right) SFRDs, while the bottom panels show the ratio of the Integral method and global SAM results with respect to our NN-based semi-numeric values.}
    \label{fig:SFRD_Comparison}
\end{figure*}

In this subsection we compare the star formation results of our ``NN-based'' model to those from the global SAM of \citetalias{Paper1}, as well as those resulting from a simpler analytic model that relies on integration of the halo mass function to predict the SFRD. Such models have frequently been used in semi-numerical simulations as a prescription for small-scale star formation \citep[e.g.][]{Visbal15a, Visbal15b, Mashian16, Munoz22, Munoz23}. By selecting distinct halo mass ranges for PopII and PopIII stars, these integrals determine the collapsed mass fraction within DM halos at redshift $z$, which is then used to calculate the SFRDs. 

We implement this analytic model (which we will refer to as the ``Integral method'' henceforth) in our large-scale 3D simulation using the same overdensity and \vbc\ fields of Figure \ref{fig:field-slices}. Within each cell, instead of calling on NN emulators to predict star formation for a given merger history, we integrate the $\delta$-dependent halo mass function (equation \ref{EQ:HMF}) over halo mass regimes corresponding to either PopII or PopIII star formation. This gives us an estimate for the cosmic mass fraction that has collapsed into DM halos, $F_{\rm coll}$, which for stellar population $i$ is given by
\begin{equation}
    F_{\mathrm{coll},i}(z) = \frac{1}{\Omega_{\rm m}\rho_{\rm c}} \int_{M_{l,i}}^{M_{u,i}} M \frac{dn}{dM}(z) \ dM .
\end{equation}
Here, $\Omega_{\rm m}$ is the cosmological density parameter for matter and $\rho_{\rm c}$ is the critical density. For PopIII stars we integrate the mass function from $M_{\rm l,i}=$ \Mcrit$(z)$ to $M_{\rm u,i}=$ 2.5\Mcrit$(z)$, and integrate over higher DM halo masses ($\leq 10^{13}$ \Msun) to estimate PopII star formation. The time derivatives for both stellar populations, $(dF_{\rm coll}/{dt})_{i}$, are then calculated numerically and multiplied by $\rho_{\rm b}f_{i}$ to give the PopII and PopIII SFRDs given some star formation efficiency, $f_{i}$. It is important to note that for the Integral method, we assume a PopII star formation efficiency of $f_{\rm II} = 0.005$, and a new PopIII stellar mass of 400 \Msun\ (approximately corresponding to a constant PopIII star formation efficiency of $f_{\rm III} = 0.0002$ for $M_{\rm halo} = 10^{6.1}$ \Msun). These values are each double those of our NN-based semi-numerical framework which, as in \citetalias{Paper1}, were chosen to more closely match the SFRD results of our full simulation.

In Figure \ref{fig:SFRD_Comparison}, we compare the average \JLW, $\rm SFRD_{III}(z)$, and $\rm SFRD_{II}(z)$ of our simulation to those resulting from the Integral method and global SAM. We do so to illustrate how the star formation evolution changes when one includes merger histories and 3D fluctuations in both matter density and feedback. From the left panels, we see that our semi-numeric and Integral method LW background intensities very closely agree for $z \lesssim 45$, only diverging at the very end of the simulation as the emulated PopII SFRD outpaces that of the Integral method. Similarly, the \JLW\ of the global SAM at $z \lesssim 30$ agrees with our full simulation to within tens of percent, but typically underestimates star formation and therefore feedback as its merger history was only well-converged out to $z = 35$.

We see from Figure \ref{fig:SFRD_Comparison} that the 3D semi-numeric $\rm SFRD_{III}$ histories trace each other to within $10\%$ at all $z \lesssim 45$. Conversely, the finite halo mass resolution of the global SAM causes a delay in star formation until a sufficient number of halos forms and both PopIII and PopII SFRDs rise to match our full model at the tens of percent level for $z\lesssim30$. The Integral method model also follows the NN-based PopII star formation history to within tens of percent at all $z\lesssim38$, albeit growing at a slightly shallower pace, yielding a final \SMD\ value that is about half the semi-numeric value. The general agreement across cosmic time indicates that the average large-scale SFRD and LW feedback are relatively unaffected by the inclusion of 3D fluctuations in analytic integration models, provided one picks reasonable bounds of integration.

\begin{figure*}
    \centering
    \includegraphics[width=\textwidth]{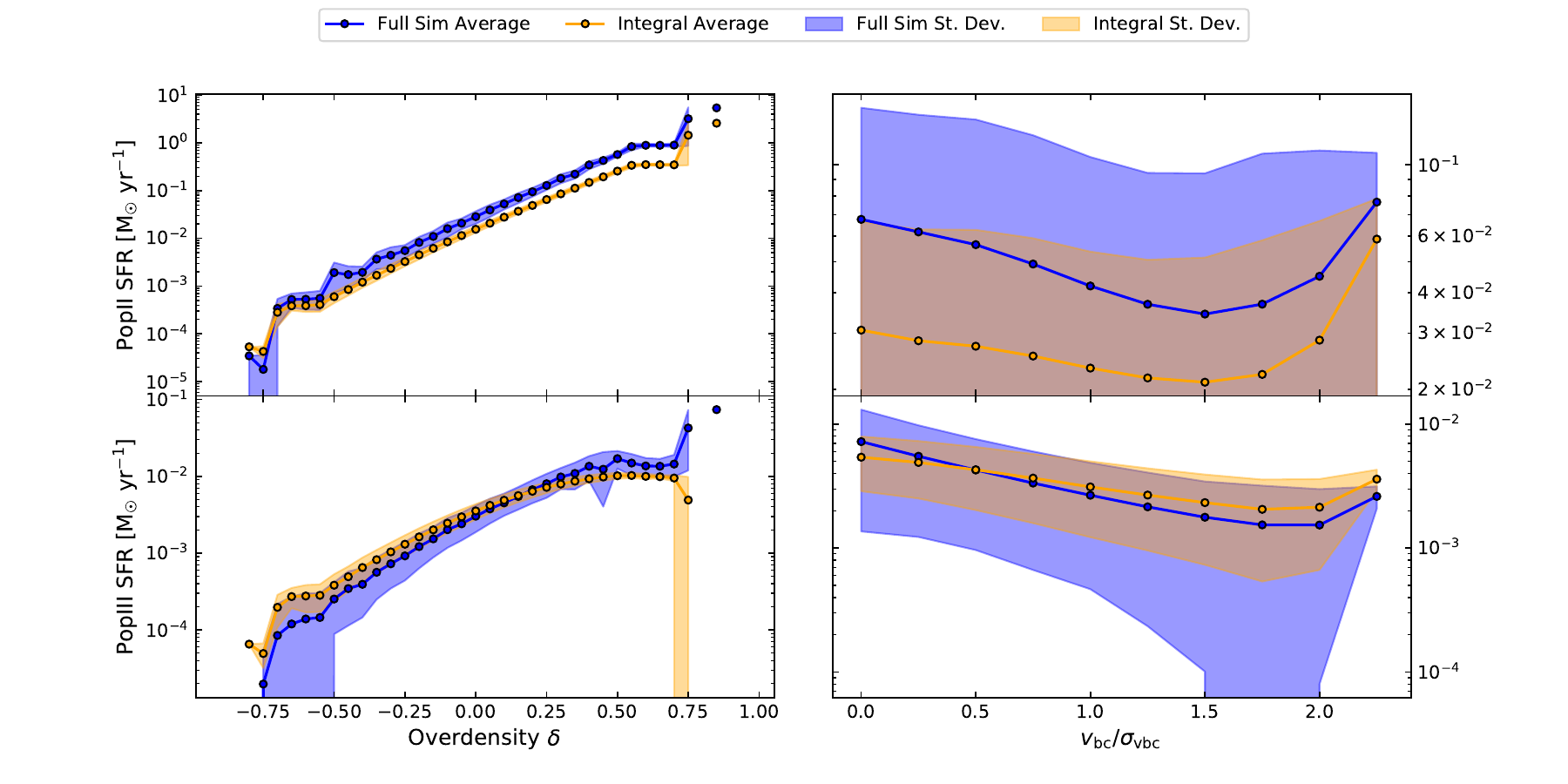}
    \caption{The average cell SFRs of our NN-based semi-numeric simulation (blue) compared to those from the Integral method (orange) at $z =$ 20. \textbf{Left Column: } Comparison of average SFRs binned by cell overdensity, \overdense, and separated by PopII (top) and PopIII (bottom) star formation. The average SFR of each bin is shown by the outlined circles, and the shaded regions represent a 1$\sigma$ standard deviation about the average. Note that the isolated points at the highest overdensities are separated from the other data points by an overdensity bin containing zero cells, as such high \overdense\ values are exceedingly rare. \textbf{Right Column: } The same as the left column, but with SFR binned by cell \vbc\ in terms of its rms value at Recombination, \sigvbc. Values for \overdense\ and \vbc\ are those generated by 21cmvFast, and correspond to values at $z = 15$ and $z = 1060$, respectively \citep[][also see Section \ref{21cmvFast}]{Munoz19}.}
    \label{fig:SFR_by_delta_vbc}
\end{figure*}

\subsubsection{Results by Cell Environment} \label{subsub:cell_environment}
We now shift our focus to compare the 3D models of this work; in Figure \ref{fig:SFR_by_delta_vbc} we compare the average SFR($z=20$) of cells in our full simulation to those of the Integral method in terms of their \overdense\ and \vbc. This particular redshift is not motivated by astrophysics, rather we chose an intermediate redshift after roughly all cells have begun star formation, but before the end of the simulation when our DM halo merger trees converge into single final halo masses.

From the left column of Figure \ref{fig:SFR_by_delta_vbc}, we see that the overdensity of a cell drastically affects its SFR. The average PopIII SFR in the most overdense cells of our semi-numeric framework is over four orders of magnitude higher than the SFR of our most underdense cells. Alternatively, the PopIII SFRs of the Integral method span just over two orders of magnitude from the most to least overdense cells, indicating that differences in \overdense\ more heavily affect SFRs in our semi-numeric simulation. Similarly, the PopII SFRs in the top left span $\sim$5.5 and $\sim$4.8 orders of magnitude for the semi-numeric and Integral methods, respectively. 

\begin{figure*}
    \centering
    \includegraphics[width=\textwidth]{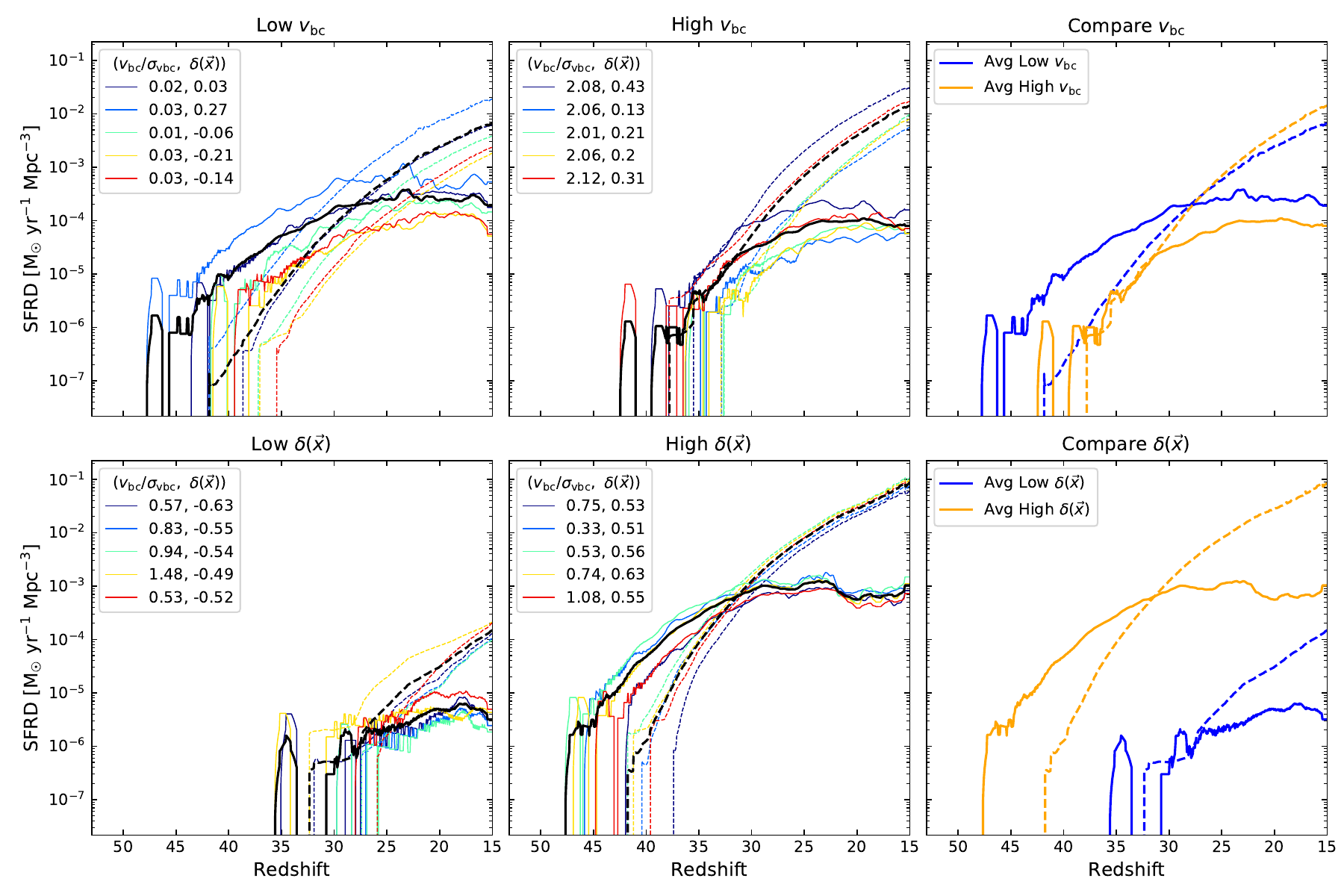}
    \caption{A comparison of the PopIII (solid) and PopII (dashed) SFRDs of five randomly sampled cells from our simulation for different values of \sigvbc\ and \overdense, as noted by the values in the legends of the left two columns. \textbf{Top Row:} SFRD comparison of five sample cells with relatively low (left) and high (middle) relative baryon-DM streaming velocities. We average the SFRDs in each panel by stellar population (black curves), and compare these to one another in the right panel to emphasize the impact of different \vbc\ on the SFRD. \textbf{Bottom Row: } Same as the top row, but for cells with relatively low (left) and high (middle) cosmic overdensities. The SFRD averages of these two panels are again compared in the right panel.}
    \label{fig:SFRD_Samples}
\end{figure*}

Looking to the right column of Figure \ref{fig:SFR_by_delta_vbc}, we find that the \vbc\ of a cell has a subdominant effect on the SFR when compared to the \overdense. While the critical mass for PopIII star formation depends on the local streaming velocity, the average SFRs only span a factor of $\sim$5 and $\sim$2.6 for our semi-numeric and Integral methods, respectively. The PopII SFRs are not as directly dependent on the \vbc, and so differences in merger history, \JLW, and the available material for star formation diminishes their dynamic range to a factor of $\sim$2.5 in both models. Further, both models and stellar populations reach a minimum SFR at streaming velocities of \sigvbc $=$ 1.5-1.75, despite the intuitive expectation that the highest \vbc\ cells would have the most suppressed star formation. By analyzing the average overdensity of cells in each \vbc\ bin, we find that it hovers around zero at \sigvbc\ $\lesssim 1.75$ and begins to increase sharply for higher streaming velocities. We therefore attribute this behavior to the fact that more overdense cells tend to have higher \vbc, meaning the highest velocity bin also contains more overdense cells which dominate the SFR, causing the increase at high \sigvbc. 

Overall, Figure \ref{fig:SFR_by_delta_vbc} shows that our semi-numeric simulation cover a wider dynamic range of SFRs than the Integral method, and experiences larger scatter between cells of similar \overdense\ and \vbc. The inclusion of DM halo merger histories naturally explains this diversity in cell SFRs as their stochastic mass growth induces fluctuations in star formation that are simply not captured by the Integral method. 

To further our understanding of how star formation changes in different environments, we randomly sample cells with extreme \vbc\ and \overdense\ values and compare their SFRDs in Figure \ref{fig:SFRD_Samples}. From the top row, we see that low streaming velocity cells tend to form stars earlier than cells with high \vbc. The average low and high \sigvbc\ curves in the right panel begin forming PopIII stars at $z \sim 46.5$ and $z \sim 41.5$, respectively, a difference of $\sim$49.6 Myr. On the other hand, the average PopIII SFRDs in the bottom right respectively begin at redshifts of $\sim$48 and $\sim$35 for high and low \overdense; this is a time difference of $\sim$153.4 Myr, or over three times the period between average \vbc\ curves in the top row.

Further, the extreme \vbc\ cells in the top left and middle panels tend to exhibit more scatter than the extreme \overdense\ cells in the bottom row. While differences in streaming velocity and \JLW\ certainly influence the star formation history of a cell, its overdensity typically dominates both when it begins star formation relative to other similar cells and how much star formation ensues throughout the rest of the simulation. We see an example of this in the top left panel of Figure \ref{fig:SFRD_Samples} where three cells have identical values of \vbc/\sigvbc\ $=$ 0.03, but while the blue curve has an overdensity of 0.27, the red and yellow curves are underdense. This ultimately leads to the blue curve having final SFRD values that are approximately an order of magnitude higher than the red and yellow curves, further emphasizing our finding that a higher \vbc\ has a subdominant effect on the SFR relative to the overdensity.

\begin{SCfigure}
    \centering
    \includegraphics[width=0.55\textwidth]{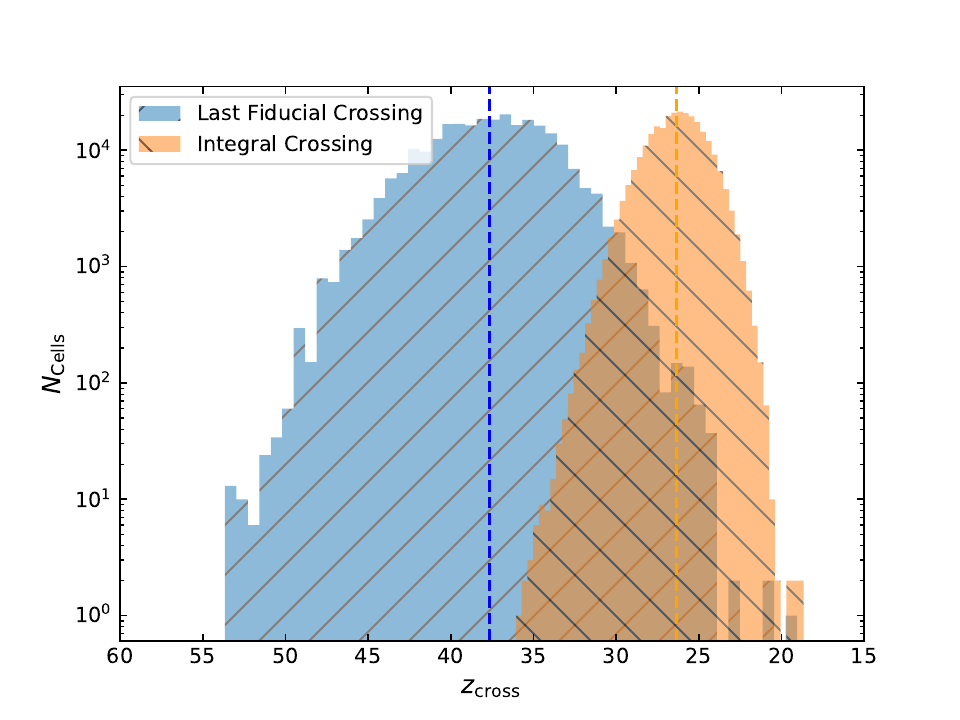}
    \caption{Histograms of \zcross, the redshift at which the PopII SFRD surpasses the PopIII SFRD, for all cells in the NN-based (blue) and Integral method (orange) simulations. Due to the noisier SFRD evolution of our NN-based model, the PopII SFRD may surpass the PopIII more than once over the course of the simulation, and so we choose to show the \emph{final} time this occurs. Conversely, the Integral method SFRDs evolve more smoothly, meaning the PopII SFRD only overtakes the PopIII at one single redshift in all cells. Weighted average redshifts of each distribution are shown by the correspondingly colored vertical dashed lines.}
    \label{fig:SFR_Cross}
\end{SCfigure}

\begin{figure*}
    \centering
    \includegraphics[width=\textwidth]{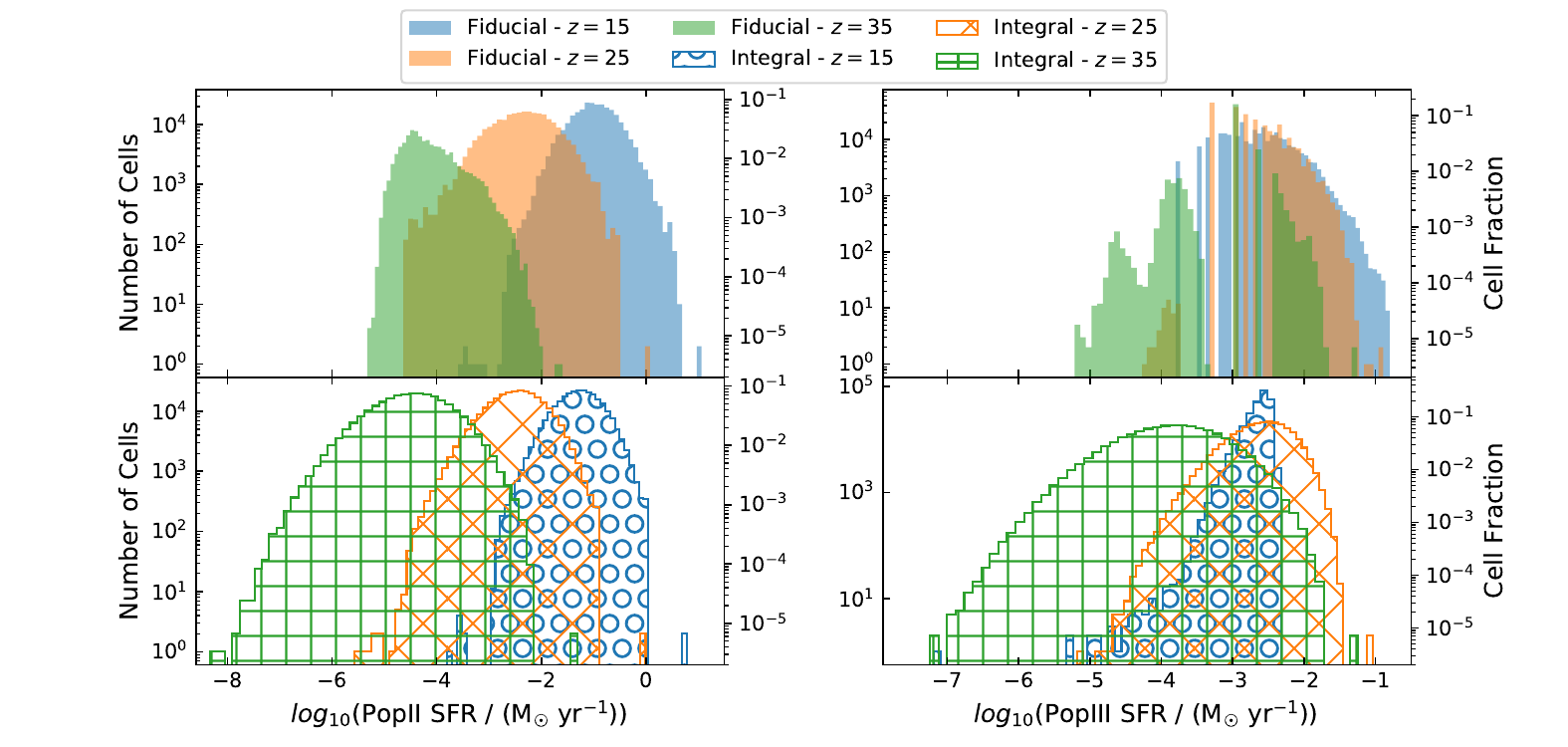}
    \caption{Histograms of cell SFRs given by our semi-numeric NN-based (top row, solid colors) and Integral method (bottom row, hatched lines) models. We show the distribution of PopII (left) and PopIII (right) SFRs in logarithmic space for three redshifts: $z =$ 15 (blue), $z =$ 25 (orange), and $z =$ 30 (green). The left axis of each panel corresponds to the number of cells in each SFR bin, while the right axes denote the fraction of total cells in the volume.}
    \label{fig:SFR_Hists}
\end{figure*}

\subsubsection{3D Star Formation Results \& Comparison} \label{subsub:SFRs_Integral}
We now utilize our model to explore when the transition between PopIII and PopII-dominated star formation occurs. The rightmost column of Figure \ref{fig:SFRD_Samples} shows that the environmental conditions of a cell can affect the redshift at which the PopII SFRD ultimately surpasses the PopIII SFRD, which we define as \zcross. From high to low \vbc\ cells, this shifts from \zcross\ $\sim$33 to \zcross\ $\sim$28. Similarly, from high to low overdensity cells, this shifts from \zcross\ $\sim$31 to \zcross\ $\sim$27.5. We extend this analysis to include all simulation cells, and plot the distributions of \zcross\ for both 3D models in Figure \ref{fig:SFR_Cross}. 

Owing to statistical variations in merger history and star formation in our full model, the resulting SFRD evolution experiences far more fluctuations than those of the Integral method. This means that for the average cell of our simulation, the PopII SFRD surpasses the PopIII multiple times, with the maximum being 21 SFRD crossings for a single cell. Further, $\sim 0.01\%$ of all cells never experience PopII domination in our simulation. We therefore plot the distribution of the final \zcross\ values from our full model in Figure \ref{fig:SFR_Cross}. This is the most statistically reliable \zcross\ since larger fluctuations in star formation occur at higher redshifts, causing PopIII SFRD spikes that quickly decrease after. Conversely, the Integral method SFRDs evolve more smoothly, and so once PopII star formation begins to dominate the total SFRD, it stays dominant for the remainder of the simulation. 

The average \zcross\ is $z = 37.63$ and $z = 26.34$ for the NN-based and Integral method simulations, respectively. This means that on average, the PopII SFRDs begin to dominate in cells with halo merger history information $\sim$48.7 Myr earlier than in the Integral method. We note that our choice of \Mcrit\ model \citep{Kulkarni21} tends to give lower values than similar contemporary models \citep[e.g.][]{Schauer21, Nebrin23}, causing \zcross\ to occur earlier as halos more easily overcome \Mcrit. While there is significant overlap in the two distributions of Figure \ref{fig:SFR_Cross}, the NN-based distribution is broader, noisier, and centered at higher redshifts, with a tail extending out to $z\sim53$. 

In Figure \ref{fig:SFR_Hists}, we compare the distributions of PopII and PopIII SFRs given by both 3D models at three sample redshifts. At $z=35$, the PopII SFR distributions for both simulations extend up to $\rm log_{10}(SFR_{II} / (M_{\odot}\ yr^{-1})) = -2$, while the Integral method extends to much lower SFR than the semi-numeric. This is because the Integral method includes contributions from all halos up to $M_{\rm halo} = 10^{13}$ \Msun\ and is not realistically bound to any one halo mass growth history. This means that fluctuations in mass growth history are lost and cell SFRs more steadily rises throughout the simulation. As time progresses, the dynamic range of PopII SFRs decrease and the distributions shift to higher values, with the semi-numeric values rising more rapidly and maintaining a broader range due to its diversity in halo merger history

Looking to the right panels, we see more apparent differences between simulation methods. At $z=35$ the two methods again have similar max SFRs, with the Integral method distribution extending to lower values, but the NN-based distribution is far noisier than the Integral method due in part to our PopIII star formation prescription. While the Integral method PopIII SFRs appear smoothly distributed, over $10\%$ of all cells in our semi-numeric simulation are concentrated in a spike at $\rm log_{10}(SFR_{III}\ [M_{\odot}\ yr^{-1}]) \approx -2.92$ which exactly corresponds to \MIII\ $=$ 1 at $z=35$, or cells in which 200 \Msun\ of new PopIII stellar mass is being emulated as a single DM halo within them overcomes \Mcrit. Further, the SFR bins at $\rm log_{10}(SFR_{III}\ [M_{\odot}\ yr^{-1}]) >$ -2.92 correspond to increasing integer \MIII\ values for cells in which multiple halos surpass \Mcrit\ over the same redshift step. Many cells, however, have $\rm log_{10}(SFR\  [M_{\odot}\ yr^{-1}]) \lesssim -3.4$ which corresponds to a fractional $\bar M_{\rm III,200}(z)$ value. As mentioned in \ref{sub:PopIII-emulation}, we do not round $\bar M_{\rm III,200}(z)$ down if it is less than one to preserve the onset timing of PopIII star formation in each cell and ensure accurate NN emulation. As time progresses, however, small PopIII star formation events give way to larger bursts, and so the fractional $\bar M_{\rm III,200}(z)$ distribution is greatly diminished at redshift 25, and disappears completely by $z = 15$. Meanwhile, the Integral method maintains a singly peaked distribution of PopIII SFRs throughout the simulation due to its smooth star formation prescription. It only experiences a notable shift between redshifts 25 and 15 as a result of increased PopII SFR and LW feedback, causing a suppression of PopIII star formation and skewing the distribution to the left.

\begin{figure*}
    \centering
    \includegraphics[width=\textwidth]{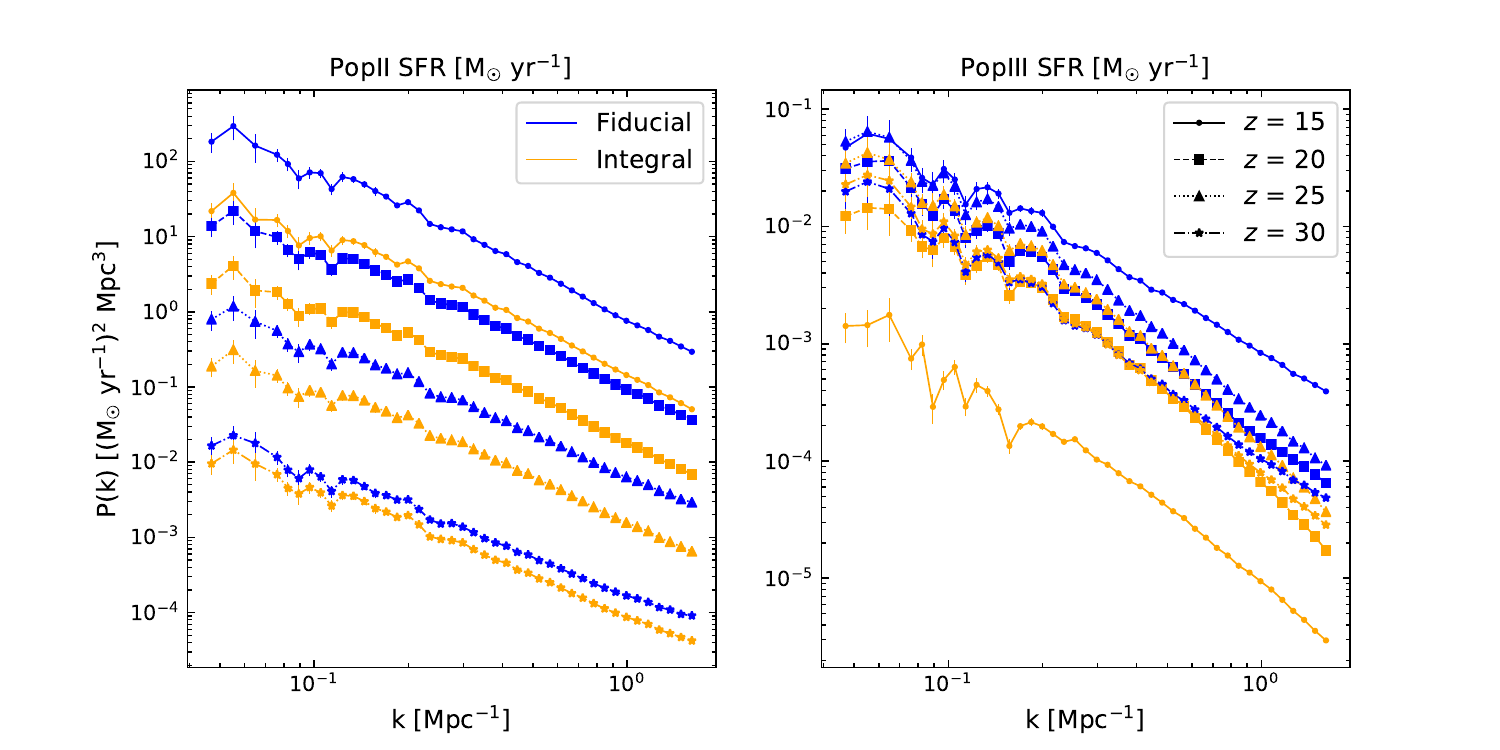}
    \caption{A comparison of the PopII (left) and PopIII (right) SFR power spectra at various redshifts, resulting from our semi-numeric NN-based (blue) and Integral method (orange) simulations. We show power spectra at $z$ = 15 (solid line, circles), 20 (dashed line, squares), 25 (dotted line, triangles), and 30 (dot-dashed line, stars).}
    \label{fig:Power_Spectra}
\end{figure*}

As a final means of comparison between semi-numerical models, we show power spectra of their PopII and PopIII SFRs at various redshifts in Figure \ref{fig:Power_Spectra}. Throughout the simulation the PopII SFR power spectra of both models steadily rise with time, with the semi-numeric simulation overtaking the Integral method at $z \sim 30$, following the PopII SFRD trend in Figure \ref{fig:SFRD_Comparison}. Similarly, as the LW background of the full simulation overtakes that of the Integral method ($z \sim 25$), the PopIII SFR power spectrum of the Integral model begins to decrease while the semi-numeric spectrum continues growing, leading to the large disparity between the two models at $z = 15$.

Beyond the overall growth, the most obvious distinction between the two methods is the slight flattening of the NN-based power spectra at large $k$, regardless of redshift or stellar population. This difference is more pronounced for PopIII, and the clearest example is seen by the $z=30$ PopIII power spectra where the two methods exhibit similar power for modes of $k \lesssim 0.4\ \rm Mpc^{-1}$, but the full simulation begins to level off with increasing $k$ while the Integral method continues decreasing. The SFR power spectra trace the underlying matter power spectrum on large distance scales ($k \lesssim 0.3\ \rm Mpc^{-1}$), and so the similarity between models on these scales is due to the SFR being dominated by large-scale density field. The flattening relative to the matter power spectrum on small scales is also present in both models, though the effect is stronger in the NN-based simulation as it includes DM halo merger trees, which further boosts fluctuations between cells. We also find that this effect decreases with time as the models more closely trace the small-scale matter power spectrum, leading to a reduction in the flattening seen on such scales in Figure \ref{fig:Power_Spectra}. 

We again note that our overdensity binning procedure resulted in 400 unique DM halo merger histories with which we populated our simulation cells; this lead to $\sim 2.58\%$ of all $64^{3}$ cells neighboring another cell with the identical merger history (Section \ref{sub:NN_Training}). To ensure that our results were insensitive to such adjacent copies, we tested a simulation using 100 and 200 overdensity bins. We found that the power spectra converged between 200 and 400 bins to within $\sim$4\% on average, hence our fiducial choice of 400 bins.

\section{Summary and Conclusions} \label{Conclusions}

A challenge in simulating the first stars and galaxies is the vast range of spatial scales that must be taken into account. It is computationally prohibitive to simulate the volumes necessary for a self-consistent treatment of LW feedback ($\gtrsim 100~{\rm Mpc}$) while resolving the dark matter minihalos which host the first stars. Most previous semi-numerical models of the large-scale distribution of the first stars and galaxies have addressed this by employing simple sub-grid analytic prescriptions for star formation  (e.g., integrating the halo mass function). 

In this paper, we developed a new semi-numerical framework that self-consistently simulates LW feedback on $\gtrsim 100~{\rm Mpc}$ scales, while simultaneously including detailed merger histories of dark matter halos that resolve the low-mass minihalos relevant for PopIII star formation. We accomplish this by creating NNs to emulate an MC merger tree-based SAM of the first stars and galaxies \citepalias{Paper1}. The NN emulators rapidly determine the PopIII and PopII star formation history of $\rm (3\ Mpc)^{3}$ cells within our simulation by taking into account the unique merger histories and environments of the cells (e.g., \JLW, \overdense , and \vbc). 

Once trained on a representative set of SAM data, we find that the emulator can accurately reproduce the star formation history of a cell to within $< 10 \%$ on average (compared to what would have been computed directly by the SAM), but with a speedup of more than two orders of magnitude. This computational acceleration enables us to self-consistently follow star formation across large distances while including a sophisticated model within much smaller regions. We note that currently the training stage is relatively computationally expensive in our model ($\sim$32,000 CPU hours), but we have not worked substantially to optimize this aspect of the framework here. 

We utilized our framework to simulate PopIII and PopII star formation in a $(192~{\rm Mpc})^3$ Mpc box from $z=60-15$ and compared the results with the global SAM in \citetalias{Paper1} and with a second 3D semi-numerical simulation where star formation was computed with analytic halo mass function integration. The latter (``Integral method'') approach is intended to roughly mimic the prescriptions employed in most previous works \citep[e.g.][]{Fialkov12, Fialkov13, Visbal15a, Visbal15b, Munoz22, Munoz23, Cruz24}. When comparing to the global SAM, we found that the volume-averaged star formation histories of our semi-numeric framework agreed with the global model to within tens of percent at $z \lesssim 30$. This indicates that accurately including spatial fluctuations does not drastically modify the globally-averaged SFRDs. 

When comparing our NN-based star formation emulation with the Integral method, we found several important differences. First, in individual simulation cells, the integral method yields smoother stellar mass growth histories that do not accurately capture the bursty star formation of DM halo merger trees. We also find that the Integral method has a smaller scatter in $z_{\rm cross}$ (the transition redshift between PopIII/PopII dominated star formation) and that this transition occurs far later than in our NN-based model. Additionally, when looking at the spatial clustering of star formation rates, we see that the diversity of halo merger histories in our NN-based model leads to substantially larger values of power spectra of PopIII and PopII star formation rates at high wavemodes.

We note that each of our NN models is trained for one specific set of merger trees within a $\rm (3\ Mpc)^{3}$ simulation cell. In our framework, it would be computationally prohibitive to generate a different NN for each of the $64^3$ cells in our simulation. However, we have generated enough unique merger histories to achieve convergence in the quantities we analyzed (e.g., the global SFRD and power spectra of PopIII/PopII SFRs). It is important to note that some quantities take more merger histories than others to reach convergence (e.g., the SFR power spectra compared to the mean SFRDs). Fortunately, all properties of the simulation we examined converged with a feasible number of merger histories.

In future work, we intend to explore a variety of machine learning architectures and data sampling strategies to accelerate the NN training stage of our simulation. Currently, our trained model runs on one CPU; we intend to parallelize the simulation to allow much larger realizations volumes in a reasonable amount of time (current realizations take $\sim$96 CPU-hours to run). We also plan to boost the accuracy of our small-scale model by including more sophisticated models for star formation, e.g. including supernovae, more accurate models for \tdelay, and varying the PopIII initial mass function. It will also be straightforward to train our model with N-body based SAMs \citep[e.g.][]{Visbal20, Hartwig22} that take into account small-scale 3D information such as fluctuations in the LW background within cells and external metal enrichment. Studies of alternate cosmological models may also be performed with this framework, but will require re-training of the NNs for each set of assumed parameters as the model heavily depends on cosmology, so this is left to a future simulation versions with a fully optimized architecture and pipeline. Finally, we will utilize our framework to make improved predictions for the cosmological 21-cm signal at cosmic dawn observable with instruments such as HERA and SKA.

\acknowledgments

We acknowledge support from NSF grant AST-2009309 and NASA ATP grant 80NNSSC22K0629. The majority of our computations were carried out at the Ohio Supercomputer Center.

\bibliographystyle{JHEP}
\bibliography{main}

\providecommand{\href}[2]{#2}\begingroup\raggedright\begin{thebibliography}{10}

\bibitem{Abel02}
T.~{Abel}, G.L.~{Bryan} and M.L.~{Norman}, \emph{{The Formation of the First Star in the Universe}}, \href{https://doi.org/10.1126/science.295.5552.93}{\emph{Science} {\bfseries 295} (2002) 93} [\href{https://arxiv.org/abs/astro-ph/0112088}{{\ttfamily astro-ph/0112088}}].

\bibitem{Bryan14-ENZO}
G.L.~{Bryan}, M.L.~{Norman}, B.W.~{O'Shea}, T.~{Abel}, J.H.~{Wise}, M.J.~{Turk} et~al., \emph{{ENZO: An Adaptive Mesh Refinement Code for Astrophysics}}, \href{https://doi.org/10.1088/0067-0049/211/2/19}{\emph{ApJs} {\bfseries 211} (2014) 19} [\href{https://arxiv.org/abs/1307.2265}{{\ttfamily 1307.2265}}].

\bibitem{Hirano14}
S.~{Hirano}, T.~{Hosokawa}, N.~{Yoshida}, H.~{Umeda}, K.~{Omukai}, G.~{Chiaki} et~al., \emph{{One Hundred First Stars: Protostellar Evolution and the Final Masses}}, \href{https://doi.org/10.1088/0004-637X/781/2/60}{\emph{ApJ} {\bfseries 781} (2014) 60} [\href{https://arxiv.org/abs/1308.4456}{{\ttfamily 1308.4456}}].

\bibitem{Klessen&Glover23}
R.S.~{Klessen} and S.C.O.~{Glover}, \emph{{The first stars: formation, properties, and impact}}, \href{https://doi.org/10.48550/arXiv.2303.12500}{\emph{arXiv e-prints} (2023) arXiv:2303.12500} [\href{https://arxiv.org/abs/2303.12500}{{\ttfamily 2303.12500}}].

\bibitem{Schaerer02}
D.~{Schaerer}, \emph{{On the properties of massive Population III stars and metal-free stellar populations}}, \href{https://doi.org/10.1051/0004-6361:20011619}{\emph{AAP} {\bfseries 382} (2002) 28} [\href{https://arxiv.org/abs/astro-ph/0110697}{{\ttfamily astro-ph/0110697}}].

\bibitem{Smith15}
B.D.~{Smith}, J.H.~{Wise}, B.W.~{O'Shea}, M.L.~{Norman} and S.~{Khochfar}, \emph{{The first Population II stars formed in externally enriched mini-haloes}}, \href{https://doi.org/10.1093/mnras/stv1509}{\emph{MNRAS} {\bfseries 452} (2015) 2822} [\href{https://arxiv.org/abs/1504.07639}{{\ttfamily 1504.07639}}].

\bibitem{Chiaki13}
G.~{Chiaki}, N.~{Yoshida} and T.~{Kitayama}, \emph{{Low-mass Star Formation Triggered by Early Supernova Explosions}}, \href{https://doi.org/10.1088/0004-637X/762/1/50}{\emph{ApJ} {\bfseries 762} (2013) 50} [\href{https://arxiv.org/abs/1203.0820}{{\ttfamily 1203.0820}}].

\bibitem{Chiaki17}
G.~{Chiaki}, H.~{Susa} and S.~{Hirano}, \emph{{Formation environment of Pop II stars affected by the feedbacks from Pop III stars}}, {\emph{MemSAIt} {\bfseries 88} (2017) 856}.

\bibitem{Skinner20}
D.~{Skinner} and J.H.~{Wise}, \emph{{Cradles of the first stars: self-shielding, halo masses, and multiplicity}}, \href{https://doi.org/10.1093/mnras/staa139}{\emph{MNRAS} {\bfseries 492} (2020) 4386} [\href{https://arxiv.org/abs/2001.04480}{{\ttfamily 2001.04480}}].

\bibitem{Naidu22}
R.P.~{Naidu}, P.A.~{Oesch}, P.~{van Dokkum}, E.J.~{Nelson}, K.A.~{Suess}, G.~{Brammer} et~al., \emph{{Two Remarkably Luminous Galaxy Candidates at z {\ensuremath{\approx}} 10-12 Revealed by JWST}}, \href{https://doi.org/10.3847/2041-8213/ac9b22}{\emph{ApJl} {\bfseries 940} (2022) L14} [\href{https://arxiv.org/abs/2207.09434}{{\ttfamily 2207.09434}}].

\bibitem{Labbe23}
I.~{Labb{\'e}}, P.~{van Dokkum}, E.~{Nelson}, R.~{Bezanson}, K.A.~{Suess}, J.~{Leja} et~al., \emph{{A population of red candidate massive galaxies 600 Myr after the Big Bang}}, \href{https://doi.org/10.1038/s41586-023-05786-2}{\emph{Nature} {\bfseries 616} (2023) 266} [\href{https://arxiv.org/abs/2207.12446}{{\ttfamily 2207.12446}}].

\bibitem{Finkelstein23}
S.L.~{Finkelstein}, M.B.~{Bagley}, H.C.~{Ferguson}, S.M.~{Wilkins}, J.S.~{Kartaltepe}, C.~{Papovich} et~al., \emph{{CEERS Key Paper. I. An Early Look into the First 500 Myr of Galaxy Formation with JWST}}, \href{https://doi.org/10.3847/2041-8213/acade4}{\emph{ApJl} {\bfseries 946} (2023) L13} [\href{https://arxiv.org/abs/2211.05792}{{\ttfamily 2211.05792}}].

\bibitem{Frebel15}
A.~{Frebel} and J.E.~{Norris}, \emph{{Near-Field Cosmology with Extremely Metal-Poor Stars}}, \href{https://doi.org/10.1146/annurev-astro-082214-122423}{\emph{ARAA} {\bfseries 53} (2015) 631} [\href{https://arxiv.org/abs/1501.06921}{{\ttfamily 1501.06921}}].

\bibitem{Hartwig15}
T.~{Hartwig}, V.~{Bromm}, R.S.~{Klessen} and S.C.O.~{Glover}, \emph{{Constraining the primordial initial mass function with stellar archaeology}}, \href{https://doi.org/10.1093/mnras/stu2740}{\emph{MNRAS} {\bfseries 447} (2015) 3892} [\href{https://arxiv.org/abs/1411.1238}{{\ttfamily 1411.1238}}].

\bibitem{deBennassuti17}
M.~{de Bennassuti}, S.~{Salvadori}, R.~{Schneider}, R.~{Valiante} and K.~{Omukai}, \emph{{Limits on Population III star formation with the most iron-poor stars}}, \href{https://doi.org/10.1093/mnras/stw2687}{\emph{MNRAS} {\bfseries 465} (2017) 926} [\href{https://arxiv.org/abs/1610.05777}{{\ttfamily 1610.05777}}].

\bibitem{Graziani17}
L.~{Graziani}, M.~{de Bennassuti}, R.~{Schneider}, D.~{Kawata} and S.~{Salvadori}, \emph{{The history of the dark and luminous side of Milky Way-like progenitors}}, \href{https://doi.org/10.1093/mnras/stx900}{\emph{MNRAS} {\bfseries 469} (2017) 1101} [\href{https://arxiv.org/abs/1704.02983}{{\ttfamily 1704.02983}}].

\bibitem{Griffen18}
B.F.~{Griffen}, G.A.~{Dooley}, A.P.~{Ji}, B.W.~{O'Shea}, F.A.~{G{\'o}mez} and A.~{Frebel}, \emph{{Tracing the first stars and galaxies of the Milky Way}}, \href{https://doi.org/10.1093/mnras/stx2749}{\emph{MNRAS} {\bfseries 474} (2018) 443} [\href{https://arxiv.org/abs/1611.00759}{{\ttfamily 1611.00759}}].

\bibitem{Magg18}
M.~{Magg}, T.~{Hartwig}, B.~{Agarwal}, A.~{Frebel}, S.C.O.~{Glover}, B.F.~{Griffen} et~al., \emph{{Predicting the locations of possible long-lived low-mass first stars: importance of satellite dwarf galaxies}}, \href{https://doi.org/10.1093/mnras/stx2729}{\emph{MNRAS} {\bfseries 473} (2018) 5308} [\href{https://arxiv.org/abs/1706.07054}{{\ttfamily 1706.07054}}].

\bibitem{Hartwig18b}
T.~{Hartwig}, V.~{Bromm} and A.~{Loeb}, \emph{{Detection strategies for the first supernovae with JWST}}, \href{https://doi.org/10.1093/mnras/sty1576}{\emph{MNRAS} {\bfseries 479} (2018) 2202} [\href{https://arxiv.org/abs/1711.05742}{{\ttfamily 1711.05742}}].

\bibitem{Lorenzo22}
M.~{Lorenzo}, M.~{Garcia}, F.~{Najarro}, A.~{Herrero}, M.~{Cervi{\~n}o} and N.~{Castro}, \emph{{A new reference catalogue for the very metal-poor Universe: +150 OB stars in Sextans A}}, \href{https://doi.org/10.1093/mnras/stac2050}{\emph{MNRAS} {\bfseries 516} (2022) 4164} [\href{https://arxiv.org/abs/2207.09700}{{\ttfamily 2207.09700}}].

\bibitem{Haiman03}
Z.~{Haiman} and G.P.~{Holder}, \emph{{The Reionization History at High Redshifts. I. Physical Models and New Constraints from Cosmic Microwave Background Polarization}}, \href{https://doi.org/10.1086/377337}{\emph{ApJ} {\bfseries 595} (2003) 1} [\href{https://arxiv.org/abs/astro-ph/0302403}{{\ttfamily astro-ph/0302403}}].

\bibitem{Shull08}
J.M.~{Shull} and A.~{Venkatesan}, \emph{{Constraints on First-Light Ionizing Sources from Optical Depth of the Cosmic Microwave Background}}, \href{https://doi.org/10.1086/590898}{\emph{ApJ} {\bfseries 685} (2008) 1} [\href{https://arxiv.org/abs/0806.0392}{{\ttfamily 0806.0392}}].

\bibitem{Ahn12}
K.~{Ahn}, I.T.~{Iliev}, P.R.~{Shapiro}, G.~{Mellema}, J.~{Koda} and Y.~{Mao}, \emph{{Detecting the Rise and Fall of the First Stars by Their Impact on Cosmic Reionization}}, \href{https://doi.org/10.1088/2041-8205/756/1/L16}{\emph{ApJl} {\bfseries 756} (2012) L16} [\href{https://arxiv.org/abs/1206.5007}{{\ttfamily 1206.5007}}].

\bibitem{Visbal15b}
E.~{Visbal}, Z.~{Haiman} and G.L.~{Bryan}, \emph{{Limits on Population III star formation in minihaloes implied by Planck}}, \href{https://doi.org/10.1093/mnras/stv1941}{\emph{MNRAS} {\bfseries 453} (2015) 4456} [\href{https://arxiv.org/abs/1505.06359}{{\ttfamily 1505.06359}}].

\bibitem{Miranda17}
V.~{Miranda}, A.~{Lidz}, C.H.~{Heinrich} and W.~{Hu}, \emph{{CMB signatures of metal-free star formation and Planck 2015 polarization data}}, \href{https://doi.org/10.1093/mnras/stx306}{\emph{MNRAS} {\bfseries 467} (2017) 4050} [\href{https://arxiv.org/abs/1610.00691}{{\ttfamily 1610.00691}}].

\bibitem{Visbal15a}
E.~{Visbal}, Z.~{Haiman} and G.L.~{Bryan}, \emph{{Looking for Population III stars with He II line intensity mapping}}, \href{https://doi.org/10.1093/mnras/stv785}{\emph{MNRAS} {\bfseries 450} (2015) 2506} [\href{https://arxiv.org/abs/1501.03177}{{\ttfamily 1501.03177}}].

\bibitem{Parsons22-HeII}
J.~{Parsons}, L.~{Mas-Ribas}, G.~{Sun}, T.-C.~{Chang}, M.O.~{Gonzalez} and R.H.~{Mebane}, \emph{{Probing Population III Initial Mass Functions with He II/H{\ensuremath{\alpha}} Intensity Mapping}}, \href{https://doi.org/10.3847/1538-4357/ac746b}{\emph{ApJ} {\bfseries 933} (2022) 141} [\href{https://arxiv.org/abs/2112.06407}{{\ttfamily 2112.06407}}].

\bibitem{Whalen13}
D.J.~{Whalen}, C.L.~{Fryer}, D.E.~{Holz}, A.~{Heger}, S.E.~{Woosley}, M.~{Stiavelli} et~al., \emph{{Seeing the First Supernovae at the Edge of the Universe with JWST}}, \href{https://doi.org/10.1088/2041-8205/762/1/L6}{\emph{ApJl} {\bfseries 762} (2013) L6} [\href{https://arxiv.org/abs/1209.3457}{{\ttfamily 1209.3457}}].

\bibitem{Hartwig18a}
T.~{Hartwig}, N.~{Yoshida}, M.~{Magg}, A.~{Frebel}, S.C.O.~{Glover}, F.A.~{G{\'o}mez} et~al., \emph{{Descendants of the first stars: the distinct chemical signature of second-generation stars}}, \href{https://doi.org/10.1093/mnras/sty1176}{\emph{MNRAS} {\bfseries 478} (2018) 1795} [\href{https://arxiv.org/abs/1801.05044}{{\ttfamily 1801.05044}}].

\bibitem{Bromm07}
V.~{Bromm} and A.~{Loeb}, \emph{{GRB Cosmology: Probing the Early Universe}},  in \emph{Supernova 1987A: 20 Years After: Supernovae and Gamma-Ray Bursters}, S.~{Immler}, K.~{Weiler} and R.~{McCray}, eds., vol.~937 of \emph{American Institute of Physics Conference Series}, pp.~532--541, Oct., 2007, \href{https://doi.org/10.1063/1.3682957}{DOI}.

\bibitem{Burlon16}
D.~{Burlon}, T.~{Murphy}, G.~{Ghirlanda}, P.J.~{Hancock}, R.~{Parry} and R.~{Salvaterra}, \emph{{Gamma-ray bursts from massive Population-III stars: clues from the radio band}}, \href{https://doi.org/10.1093/mnras/stw905}{\emph{MNRAS} {\bfseries 459} (2016) 3356} [\href{https://arxiv.org/abs/1604.03946}{{\ttfamily 1604.03946}}].

\bibitem{Kinugawa19}
T.~{Kinugawa}, Y.~{Harikane} and K.~{Asano}, \emph{{Long Gamma-Ray Burst Rate at Very High Redshift}}, \href{https://doi.org/10.3847/1538-4357/ab2188}{\emph{ApJ} {\bfseries 878} (2019) 128} [\href{https://arxiv.org/abs/1901.03516}{{\ttfamily 1901.03516}}].

\bibitem{Pritchard12}
J.R.~{Pritchard} and A.~{Loeb}, \emph{{21 cm cosmology in the 21st century}}, \href{https://doi.org/10.1088/0034-4885/75/8/086901}{\emph{Reports on Progress in Physics} {\bfseries 75} (2012) 086901} [\href{https://arxiv.org/abs/1109.6012}{{\ttfamily 1109.6012}}].

\bibitem{Bowman18-EDGES}
J.D.~{Bowman}, A.E.E.~{Rogers}, R.A.~{Monsalve}, T.J.~{Mozdzen} and N.~{Mahesh}, \emph{{An absorption profile centred at 78 megahertz in the sky-averaged spectrum}}, \href{https://doi.org/10.1038/nature25792}{\emph{Nature} {\bfseries 555} (2018) 67} [\href{https://arxiv.org/abs/1810.05912}{{\ttfamily 1810.05912}}].

\bibitem{Price18-LEDA}
D.C.~{Price}, L.J.~{Greenhill}, A.~{Fialkov}, G.~{Bernardi}, H.~{Garsden}, B.R.~{Barsdell} et~al., \emph{{Design and characterization of the Large-aperture Experiment to Detect the Dark Age (LEDA) radiometer systems}}, \href{https://doi.org/10.1093/mnras/sty1244}{\emph{MNRAS} {\bfseries 478} (2018) 4193} [\href{https://arxiv.org/abs/1709.09313}{{\ttfamily 1709.09313}}].

\bibitem{DeBoer17-HERA}
D.R.~{DeBoer}, A.R.~{Parsons}, J.E.~{Aguirre}, P.~{Alexander}, Z.S.~{Ali}, A.P.~{Beardsley} et~al., \emph{{Hydrogen Epoch of Reionization Array (HERA)}}, \href{https://doi.org/10.1088/1538-3873/129/974/045001}{\emph{PASP} {\bfseries 129} (2017) 045001} [\href{https://arxiv.org/abs/1606.07473}{{\ttfamily 1606.07473}}].

\bibitem{Mellema13-SKA}
G.~{Mellema}, L.V.E.~{Koopmans}, F.A.~{Abdalla}, G.~{Bernardi}, B.~{Ciardi}, S.~{Daiboo} et~al., \emph{{Reionization and the Cosmic Dawn with the Square Kilometre Array}}, \href{https://doi.org/10.1007/s10686-013-9334-5}{\emph{Experimental Astronomy} {\bfseries 36} (2013) 235} [\href{https://arxiv.org/abs/1210.0197}{{\ttfamily 1210.0197}}].

\bibitem{Greif15}
T.H.~{Greif}, \emph{{The numerical frontier of the high-redshift Universe}}, \href{https://doi.org/10.1186/s40668-014-0006-2}{\emph{Computational Astrophysics and Cosmology} {\bfseries 2} (2015) 3} [\href{https://arxiv.org/abs/1410.3482}{{\ttfamily 1410.3482}}].

\bibitem{Wise19}
J.H.~{Wise}, \emph{{An Introductory Review on Cosmic Reionization}}, {\emph{arXiv e-prints} (2019) arXiv:1907.06653} [\href{https://arxiv.org/abs/1907.06653}{{\ttfamily 1907.06653}}].

\bibitem{Tseliakhovich10-vbc}
D.~{Tseliakhovich} and C.~{Hirata}, \emph{{Relative velocity of dark matter and baryonic fluids and the formation of the first structures}}, \href{https://doi.org/10.1103/PhysRevD.82.083520}{\emph{PRD} {\bfseries 82} (2010) 083520} [\href{https://arxiv.org/abs/1005.2416}{{\ttfamily 1005.2416}}].

\bibitem{Tseliakhovich11}
D.~{Tseliakhovich}, R.~{Barkana} and C.M.~{Hirata}, \emph{{Suppression and spatial variation of early galaxies and minihaloes}}, \href{https://doi.org/10.1111/j.1365-2966.2011.19541.x}{\emph{MNRAS} {\bfseries 418} (2011) 906} [\href{https://arxiv.org/abs/1012.2574}{{\ttfamily 1012.2574}}].

\bibitem{Greif11}
T.H.~{Greif}, S.D.M.~{White}, R.S.~{Klessen} and V.~{Springel}, \emph{{The Delay of Population III Star Formation by Supersonic Streaming Velocities}}, \href{https://doi.org/10.1088/0004-637X/736/2/147}{\emph{ApJ} {\bfseries 736} (2011) 147} [\href{https://arxiv.org/abs/1101.5493}{{\ttfamily 1101.5493}}].

\bibitem{Fialkov12}
A.~{Fialkov}, R.~{Barkana}, D.~{Tseliakhovich} and C.M.~{Hirata}, \emph{{Impact of the relative motion between the dark matter and baryons on the first stars: semi-analytical modelling}}, \href{https://doi.org/10.1111/j.1365-2966.2012.21318.x}{\emph{MNRAS} {\bfseries 424} (2012) 1335} [\href{https://arxiv.org/abs/1110.2111}{{\ttfamily 1110.2111}}].

\bibitem{McQuinn12}
M.~{McQuinn} and R.M.~{O'Leary}, \emph{{The Impact of the Supersonic Baryon-Dark Matter Velocity Difference on the z \raisebox{-0.5ex}\textasciitilde 20 21 cm Background}}, \href{https://doi.org/10.1088/0004-637X/760/1/3}{\emph{ApJ} {\bfseries 760} (2012) 3} [\href{https://arxiv.org/abs/1204.1345}{{\ttfamily 1204.1345}}].

\bibitem{Haiman97}
Z.~{Haiman}, M.J.~{Rees} and A.~{Loeb}, \emph{{Destruction of Molecular Hydrogen during Cosmological Reionization}}, \href{https://doi.org/10.1086/303647}{\emph{ApJ} {\bfseries 476} (1997) 458} [\href{https://arxiv.org/abs/astro-ph/9608130}{{\ttfamily astro-ph/9608130}}].

\bibitem{Ahn09}
K.~{Ahn}, P.R.~{Shapiro}, I.T.~{Iliev}, G.~{Mellema} and U.-L.~{Pen}, \emph{{The Inhomogeneous Background Of H$_{2}$-Dissociating Radiation During Cosmic Reionization}}, \href{https://doi.org/10.1088/0004-637X/695/2/1430}{\emph{ApJ} {\bfseries 695} (2009) 1430} [\href{https://arxiv.org/abs/0807.2254}{{\ttfamily 0807.2254}}].

\bibitem{Haiman00}
Z.~{Haiman}, T.~{Abel} and M.J.~{Rees}, \emph{{The Radiative Feedback of the First Cosmological Objects}}, \href{https://doi.org/10.1086/308723}{\emph{ApJ} {\bfseries 534} (2000) 11} [\href{https://arxiv.org/abs/astro-ph/9903336}{{\ttfamily astro-ph/9903336}}].

\bibitem{Machacek01}
M.E.~{Machacek}, G.L.~{Bryan} and T.~{Abel}, \emph{{Simulations of Pregalactic Structure Formation with Radiative Feedback}}, \href{https://doi.org/10.1086/319014}{\emph{ApJ} {\bfseries 548} (2001) 509} [\href{https://arxiv.org/abs/astro-ph/0007198}{{\ttfamily astro-ph/0007198}}].

\bibitem{O’Shea08}
B.W.~{O'Shea} and M.L.~{Norman}, \emph{{Population III Star Formation in a {\ensuremath{\Lambda}}CDM Universe. II. Effects of a Photodissociating Background}}, \href{https://doi.org/10.1086/524006}{\emph{ApJ} {\bfseries 673} (2008) 14} [\href{https://arxiv.org/abs/0706.4416}{{\ttfamily 0706.4416}}].

\bibitem{Haiman06}
Z.~{Haiman} and G.L.~{Bryan}, \emph{{Was Star Formation Suppressed in High-Redshift Minihalos?}}, \href{https://doi.org/10.1086/506580}{\emph{ApJ} {\bfseries 650} (2006) 7} [\href{https://arxiv.org/abs/astro-ph/0603541}{{\ttfamily astro-ph/0603541}}].

\bibitem{Wyithe07}
J.S.B.~{Wyithe} and R.~{Cen}, \emph{{The Extended Star Formation History of the First Generation of Stars and the Reionization of Cosmic Hydrogen}}, \href{https://doi.org/10.1086/511948}{\emph{ApJ} {\bfseries 659} (2007) 890} [\href{https://arxiv.org/abs/astro-ph/0602503}{{\ttfamily astro-ph/0602503}}].

\bibitem{Dave12}
R.~{Dav{\'e}}, K.~{Finlator} and B.D.~{Oppenheimer}, \emph{{An analytic model for the evolution of the stellar, gas and metal content of galaxies}}, \href{https://doi.org/10.1111/j.1365-2966.2011.20148.x}{\emph{MNRAS} {\bfseries 421} (2012) 98} [\href{https://arxiv.org/abs/1108.0426}{{\ttfamily 1108.0426}}].

\bibitem{Furlanetto17}
S.R.~{Furlanetto}, J.~{Mirocha}, R.H.~{Mebane} and G.~{Sun}, \emph{{A minimalist feedback-regulated model for galaxy formation during the epoch of reionization}}, \href{https://doi.org/10.1093/mnras/stx2132}{\emph{MNRAS} {\bfseries 472} (2017) 1576} [\href{https://arxiv.org/abs/1611.01169}{{\ttfamily 1611.01169}}].

\bibitem{Furlanetto22}
S.R.~{Furlanetto} and J.~{Mirocha}, \emph{{Bursty star formation during the Cosmic Dawn driven by delayed stellar feedback}}, \href{https://doi.org/10.1093/mnras/stac310}{\emph{MNRAS} {\bfseries 511} (2022) 3895} [\href{https://arxiv.org/abs/2109.04488}{{\ttfamily 2109.04488}}].

\bibitem{Mashian16}
N.~{Mashian}, P.A.~{Oesch} and A.~{Loeb}, \emph{{An empirical model for the galaxy luminosity and star formation rate function at high redshift}}, \href{https://doi.org/10.1093/mnras/stv2469}{\emph{MNRAS} {\bfseries 455} (2016) 2101} [\href{https://arxiv.org/abs/1507.00999}{{\ttfamily 1507.00999}}].

\bibitem{Paper1}
C.R.~{Feathers}, M.~{Kulkarni}, E.~{Visbal} and R.~{Hazlett}, \emph{{A Global Semianalytic Model of the First Stars and Galaxies Including Dark Matter Halo Merger Histories}}, \href{https://doi.org/10.3847/1538-4357/ad1688}{\emph{ApJ} {\bfseries 962} (2024) 62} [\href{https://arxiv.org/abs/2306.07371}{{\ttfamily 2306.07371}}].

\bibitem{Crosby13}
B.D.~{Crosby}, B.W.~{O'Shea}, B.D.~{Smith}, M.J.~{Turk} and O.~{Hahn}, \emph{{Population III Star Formation in Large Cosmological Volumes. I. Halo Temporal and Physical Environment}}, \href{https://doi.org/10.1088/0004-637X/773/2/108}{\emph{ApJ} {\bfseries 773} (2013) 108} [\href{https://arxiv.org/abs/1306.4679}{{\ttfamily 1306.4679}}].

\bibitem{Dayal20}
P.~{Dayal}, M.~{Volonteri}, T.R.~{Choudhury}, R.~{Schneider}, M.~{Trebitsch}, N.Y.~{Gnedin} et~al., \emph{{Reionization with galaxies and active galactic nuclei}}, \href{https://doi.org/10.1093/mnras/staa1138}{\emph{MNRAS} {\bfseries 495} (2020) 3065} [\href{https://arxiv.org/abs/2001.06021}{{\ttfamily 2001.06021}}].

\bibitem{Hegde23}
S.~{Hegde} and S.R.~{Furlanetto}, \emph{{A self-consistent semi-analytic model for Population III star formation in minihalos}}, \href{https://doi.org/10.48550/arXiv.2304.03358}{\emph{arXiv e-prints} (2023) arXiv:2304.03358} [\href{https://arxiv.org/abs/2304.03358}{{\ttfamily 2304.03358}}].

\bibitem{Ventura24}
E.M.~{Ventura}, Y.~{Qin}, S.~{Balu} and J.S.B.~{Wyithe}, \emph{{Semi-analytic modelling of Pop. III star formation and metallicity evolution - I. Impact on the UV luminosity functions at z = 9-16}}, \href{https://doi.org/10.1093/mnras/stae567}{\emph{MNRAS} {\bfseries 529} (2024) 628} [\href{https://arxiv.org/abs/2401.07396}{{\ttfamily 2401.07396}}].

\bibitem{Agarwal12}
B.~{Agarwal}, S.~{Khochfar}, J.L.~{Johnson}, E.~{Neistein}, C.~{Dalla Vecchia} and M.~{Livio}, \emph{{Ubiquitous seeding of supermassive black holes by direct collapse}}, \href{https://doi.org/10.1111/j.1365-2966.2012.21651.x}{\emph{MNRAS} {\bfseries 425} (2012) 2854} [\href{https://arxiv.org/abs/1205.6464}{{\ttfamily 1205.6464}}].

\bibitem{Kulkarni13}
G.~{Kulkarni}, E.~{Rollinde}, J.F.~{Hennawi} and E.~{Vangioni}, \emph{{Chemical Enrichment of Damped Ly{\ensuremath{\alpha}} Systems as a Direct Constraint on Population III Star Formation}}, \href{https://doi.org/10.1088/0004-637X/772/2/93}{\emph{ApJ} {\bfseries 772} (2013) 93} [\href{https://arxiv.org/abs/1301.4201}{{\ttfamily 1301.4201}}].

\bibitem{Ahn21}
K.~{Ahn} and P.R.~{Shapiro}, \emph{{Cosmic Reionization May Still Have Started Early and Ended Late: Confronting Early Onset with Cosmic Microwave Background Anisotropy and 21 cm Global Signals}}, \href{https://doi.org/10.3847/1538-4357/abf3bf}{\emph{ApJ} {\bfseries 914} (2021) 44} [\href{https://arxiv.org/abs/2011.03582}{{\ttfamily 2011.03582}}].

\bibitem{Liu20}
B.~{Liu} and V.~{Bromm}, \emph{{When did Population III star formation end?}}, \href{https://doi.org/10.1093/mnras/staa2143}{\emph{MNRAS} {\bfseries 497} (2020) 2839} [\href{https://arxiv.org/abs/2006.15260}{{\ttfamily 2006.15260}}].

\bibitem{Visbal12}
E.~{Visbal}, R.~{Barkana}, A.~{Fialkov}, D.~{Tseliakhovich} and C.M.~{Hirata}, \emph{{The signature of the first stars in atomic hydrogen at redshift 20}}, \href{https://doi.org/10.1038/nature11177}{\emph{Nature} {\bfseries 487} (2012) 70} [\href{https://arxiv.org/abs/1201.1005}{{\ttfamily 1201.1005}}].

\bibitem{Visbal20}
E.~{Visbal}, G.L.~{Bryan} and Z.~{Haiman}, \emph{{Self-consistent Semianalytic Modeling of Feedback during Primordial Star Formation and Reionization}}, \href{https://doi.org/10.3847/1538-4357/ab994e}{\emph{ApJ} {\bfseries 897} (2020) 95} [\href{https://arxiv.org/abs/2001.11118}{{\ttfamily 2001.11118}}].

\bibitem{Munoz22}
J.B.~{Mu{\~n}oz}, Y.~{Qin}, A.~{Mesinger}, S.G.~{Murray}, B.~{Greig} and C.~{Mason}, \emph{{The impact of the first galaxies on cosmic dawn and reionization}}, \href{https://doi.org/10.1093/mnras/stac185}{\emph{MNRAS} {\bfseries 511} (2022) 3657} [\href{https://arxiv.org/abs/2110.13919}{{\ttfamily 2110.13919}}].

\bibitem{Munoz23}
J.B.~{Mu{\~n}oz}, \emph{{An effective model for the cosmic-dawn 21-cm signal}}, \href{https://doi.org/10.1093/mnras/stad1512}{\emph{MNRAS} {\bfseries 523} (2023) 2587} [\href{https://arxiv.org/abs/2302.08506}{{\ttfamily 2302.08506}}].

\bibitem{Cruz24}
H.A.G.~{Cruz}, J.B.~{Munoz}, N.~{Sabti} and M.~{Kamionkowski}, \emph{{The First Billion Years in Seconds: An Effective Model for the 21-cm Signal with Population III Stars}}, {\emph{arXiv e-prints} (2024) arXiv:2407.18294} [\href{https://arxiv.org/abs/2407.18294}{{\ttfamily 2407.18294}}].

\bibitem{Fialkov13}
A.~{Fialkov}, R.~{Barkana}, E.~{Visbal}, D.~{Tseliakhovich} and C.M.~{Hirata}, \emph{{The 21-cm signature of the first stars during the Lyman-Werner feedback era}}, \href{https://doi.org/10.1093/mnras/stt650}{\emph{MNRAS} {\bfseries 432} (2013) 2909} [\href{https://arxiv.org/abs/1212.0513}{{\ttfamily 1212.0513}}].

\bibitem{Fialkov14a}
A.~{Fialkov}, R.~{Barkana}, A.~{Pinhas} and E.~{Visbal}, \emph{{Complete history of the observable 21 cm signal from the first stars during the pre-reionization era}}, \href{https://doi.org/10.1093/mnrasl/slt135}{\emph{MNRAS} {\bfseries 437} (2014) L36} [\href{https://arxiv.org/abs/1306.2354}{{\ttfamily 1306.2354}}].

\bibitem{Fialkov14b}
A.~{Fialkov} and R.~{Barkana}, \emph{{The rich complexity of 21-cm fluctuations produced by the first stars}}, \href{https://doi.org/10.1093/mnras/stu1744}{\emph{MNRAS} {\bfseries 445} (2014) 213} [\href{https://arxiv.org/abs/1409.3992}{{\ttfamily 1409.3992}}].

\bibitem{Kaur2022}
H.D.~{Kaur}, Y.~{Qin}, A.~{Mesinger}, A.~{Pallottini}, T.~{Fragos} and A.~{Basu-Zych}, \emph{{The 21-cm signal from the cosmic dawn: metallicity dependence of high-mass X-ray binaries}}, \href{https://doi.org/10.1093/mnras/stac1226}{\emph{MNRAS} {\bfseries 513} (2022) 5097} [\href{https://arxiv.org/abs/2203.10851}{{\ttfamily 2203.10851}}].

\bibitem{Reis22}
I.~{Reis}, R.~{Barkana} and A.~{Fialkov}, \emph{{Shot noise and scatter in the star formation efficiency as a source of 21-cm fluctuations}}, \href{https://doi.org/10.1093/mnras/stac411}{\emph{MNRAS} {\bfseries 511} (2022) 5265} [\href{https://arxiv.org/abs/2106.13111}{{\ttfamily 2106.13111}}].

\bibitem{Magg22}
M.~{Magg}, I.~{Reis}, A.~{Fialkov}, R.~{Barkana}, R.S.~{Klessen}, S.C.O.~{Glover} et~al., \emph{{Effect of the cosmological transition to metal-enriched star formation on the hydrogen 21-cm signal}}, \href{https://doi.org/10.1093/mnras/stac1664}{\emph{MNRAS} {\bfseries 514} (2022) 4433} [\href{https://arxiv.org/abs/2110.15948}{{\ttfamily 2110.15948}}].

\bibitem{Hartwig22}
T.~{Hartwig}, M.~{Magg}, L.-H.~{Chen}, Y.~{Tarumi}, V.~{Bromm}, S.C.O.~{Glover} et~al., \emph{{Public Release of A-SLOTH: Ancient Stars and Local Observables by Tracing Halos}}, \href{https://doi.org/10.3847/1538-4357/ac7150}{\emph{ApJ} {\bfseries 936} (2022) 45} [\href{https://arxiv.org/abs/2206.00223}{{\ttfamily 2206.00223}}].

\bibitem{Villanueva-Domingo22}
P.~{Villanueva-Domingo}, F.~{Villaescusa-Navarro}, D.~{Angl{\'e}s-Alc{\'a}zar}, S.~{Genel}, F.~{Marinacci}, D.N.~{Spergel} et~al., \emph{{Inferring Halo Masses with Graph Neural Networks}}, \href{https://doi.org/10.3847/1538-4357/ac7aa3}{\emph{ApJ} {\bfseries 935} (2022) 30} [\href{https://arxiv.org/abs/2111.08683}{{\ttfamily 2111.08683}}].

\bibitem{Bonici24}
M.~{Bonici}, L.~{Biggio}, C.~{Carbone} and L.~{Guzzo}, \emph{{Fast emulation of two-point angular statistics for photometric galaxy surveys}}, \href{https://doi.org/10.1093/mnras/stae1261}{\emph{MNRAS} {\bfseries 531} (2024) 4203} [\href{https://arxiv.org/abs/2206.14208}{{\ttfamily 2206.14208}}].

\bibitem{Behera24}
J.~{Behera}, R.~{Tojeiro} and H.G.~{Chittenden}, \emph{{Optimised neural network predictions of galaxy formation histories using semi-stochastic corrections}}, \href{https://doi.org/10.48550/arXiv.2409.16548}{\emph{arXiv e-prints} (2024) arXiv:2409.16548} [\href{https://arxiv.org/abs/2409.16548}{{\ttfamily 2409.16548}}].

\bibitem{Chittenden24}
H.G.~{Chittenden}, J.~{Behera} and R.~{Tojeiro}, \emph{{Evaluating the galaxy formation histories predicted by a neural network in pure dark matter simulations}}, \href{https://doi.org/10.48550/arXiv.2409.16079}{\emph{arXiv e-prints} (2024) arXiv:2409.16079} [\href{https://arxiv.org/abs/2409.16079}{{\ttfamily 2409.16079}}].

\bibitem{Jamieson24}
D.~{Jamieson}, Y.~{Li}, F.~{Villaescusa-Navarro}, S.~{Ho} and D.N.~{Spergel}, \emph{{Field-level Emulation of Cosmic Structure Formation with Cosmology and Redshift Dependence}}, {\emph{arXiv e-prints} (2024) arXiv:2408.07699} [\href{https://arxiv.org/abs/2408.07699}{{\ttfamily 2408.07699}}].

\bibitem{Hezaveh17}
Y.D.~{Hezaveh}, L.~{Perreault Levasseur} and P.J.~{Marshall}, \emph{{Fast automated analysis of strong gravitational lenses with convolutional neural networks}}, \href{https://doi.org/10.1038/nature23463}{\emph{Nature} {\bfseries 548} (2017) 555} [\href{https://arxiv.org/abs/1708.08842}{{\ttfamily 1708.08842}}].

\bibitem{Kern17}
N.S.~{Kern}, A.~{Liu}, A.R.~{Parsons}, A.~{Mesinger} and B.~{Greig}, \emph{{Emulating Simulations of Cosmic Dawn for 21 cm Power Spectrum Constraints on Cosmology, Reionization, and X-Ray Heating}}, \href{https://doi.org/10.3847/1538-4357/aa8bb4}{\emph{ApJ} {\bfseries 848} (2017) 23} [\href{https://arxiv.org/abs/1705.04688}{{\ttfamily 1705.04688}}].

\bibitem{Shimabukuro18}
H.~{Shimabukuro} and B.~{a Semelin}, \emph{{Analysing 21cm signal with artificial neural network}},  in \emph{Peering towards Cosmic Dawn}, V.~{Jeli{\'c}} and T.~{van der Hulst}, eds., vol.~333, pp.~39--42, May, 2018, \href{https://doi.org/10.1017/S174392131701081X}{DOI}.

\bibitem{Sikder24}
S.~{Sikder}, R.~{Barkana}, I.~{Reis} and A.~{Fialkov}, \emph{{Emulation of the cosmic dawn 21-cm power spectrum and classification of excess radio models using an artificial neural network}}, \href{https://doi.org/10.1093/mnras/stad3699}{\emph{MNRAS} {\bfseries 527} (2024) 9977} [\href{https://arxiv.org/abs/2201.08205}{{\ttfamily 2201.08205}}].

\bibitem{PlanckCollaboration}
{Planck Collaboration}, {Aghanim, N.}, {Akrami, Y.}, {Ashdown, M.}, {Aumont, J.}, {Baccigalupi, C.} et~al., \emph{Planck 2018 results - vi. cosmological parameters}, \href{https://doi.org/10.1051/0004-6361/201833910}{\emph{A\&A} {\bfseries 641} (2020) A6}.

\bibitem{Nebrin24}
O.~{Nebrin}, A.~{Smith}, K.~{Lorinc}, J.~{H{\"o}rnquist}, {\r{A}}.~{Larson}, G.~{Mellema} et~al., \emph{{Lyman-$\alpha$ feedback prevails at Cosmic Dawn: Implications for the first galaxies, stars, and star clusters}}, \href{https://doi.org/10.48550/arXiv.2409.19288}{\emph{arXiv e-prints} (2024) arXiv:2409.19288} [\href{https://arxiv.org/abs/2409.19288}{{\ttfamily 2409.19288}}].

\bibitem{Kulkarni21}
M.~{Kulkarni}, E.~{Visbal} and G.L.~{Bryan}, \emph{{The Critical Dark Matter Halo Mass for Population III Star Formation: Dependence on Lyman-Werner Radiation, Baryon-dark Matter Streaming Velocity, and Redshift}}, \href{https://doi.org/10.3847/1538-4357/ac08a3}{\emph{ApJ} {\bfseries 917} (2021) 40} [\href{https://arxiv.org/abs/2010.04169}{{\ttfamily 2010.04169}}].

\bibitem{Schauer21}
A.T.P.~{Schauer}, S.C.O.~{Glover}, R.S.~{Klessen} and P.~{Clark}, \emph{{The influence of streaming velocities and Lyman-Werner radiation on the formation of the first stars}}, \href{https://doi.org/10.1093/mnras/stab1953}{\emph{MNRAS} {\bfseries 507} (2021) 1775} [\href{https://arxiv.org/abs/2008.05663}{{\ttfamily 2008.05663}}].

\bibitem{Nebrin23}
O.~{Nebrin}, S.K.~{Giri} and G.~{Mellema}, \emph{{Starbursts in low-mass haloes at Cosmic Dawn. I. The critical halo mass for star formation}}, \href{https://doi.org/10.1093/mnras/stad1852}{\emph{MNRAS} {\bfseries 524} (2023) 2290} [\href{https://arxiv.org/abs/2303.08024}{{\ttfamily 2303.08024}}].

\bibitem{Press74}
W.H.~{Press} and P.~{Schechter}, \emph{{Formation of Galaxies and Clusters of Galaxies by Self-Similar Gravitational Condensation}}, \href{https://doi.org/10.1086/152650}{\emph{ApJ} {\bfseries 187} (1974) 425}.

\bibitem{Bond91-EPSformalism}
J.R.~{Bond}, S.~{Cole}, G.~{Efstathiou} and N.~{Kaiser}, \emph{{Excursion Set Mass Functions for Hierarchical Gaussian Fluctuations}}, \href{https://doi.org/10.1086/170520}{\emph{ApJ} {\bfseries 379} (1991) 440}.

\bibitem{Lacey93}
C.~{Lacey} and S.~{Cole}, \emph{{Merger rates in hierarchical models of galaxy formation}}, \href{https://doi.org/10.1093/mnras/262.3.627}{\emph{MNRAS} {\bfseries 262} (1993) 627}.

\bibitem{Barkana04}
R.~{Barkana} and A.~{Loeb}, \emph{{Unusually Large Fluctuations in the Statistics of Galaxy Formation at High Redshift}}, \href{https://doi.org/10.1086/421079}{\emph{ApJ} {\bfseries 609} (2004) 474} [\href{https://arxiv.org/abs/astro-ph/0310338}{{\ttfamily astro-ph/0310338}}].

\bibitem{ShethTormen}
R.K.~{Sheth} and G.~{Tormen}, \emph{{Large-scale bias and the peak background split}}, \href{https://doi.org/10.1046/j.1365-8711.1999.02692.x}{\emph{MNRAS} {\bfseries 308} (1999) 119} [\href{https://arxiv.org/abs/astro-ph/9901122}{{\ttfamily astro-ph/9901122}}].

\bibitem{Moriwaki23}
K.~{Moriwaki}, T.~{Nishimichi} and N.~{Yoshida}, \emph{{Machine learning for observational cosmology}}, \href{https://doi.org/10.1088/1361-6633/acd2ea}{\emph{Reports on Progress in Physics} {\bfseries 86} (2023) 076901} [\href{https://arxiv.org/abs/2303.15794}{{\ttfamily 2303.15794}}].

\bibitem{Jeon14}
M.~{Jeon}, A.H.~{Pawlik}, V.~{Bromm} and M.~{Milosavljevi{\'c}}, \emph{{Recovery from Population III supernova explosions and the onset of second-generation star formation}}, \href{https://doi.org/10.1093/mnras/stu1980}{\emph{MNRAS} {\bfseries 444} (2014) 3288} [\href{https://arxiv.org/abs/1407.0034}{{\ttfamily 1407.0034}}].

\bibitem{Munoz19}
J.B.~{Mu{\~n}oz}, \emph{{Robust velocity-induced acoustic oscillations at cosmic dawn}}, \href{https://doi.org/10.1103/PhysRevD.100.063538}{\emph{PRD} {\bfseries 100} (2019) 063538} [\href{https://arxiv.org/abs/1904.07881}{{\ttfamily 1904.07881}}].

\bibitem{Mesinger11-21cmFAST}
A.~{Mesinger}, S.~{Furlanetto} and R.~{Cen}, \emph{{21CMFAST: a fast, seminumerical simulation of the high-redshift 21-cm signal}}, \href{https://doi.org/10.1111/j.1365-2966.2010.17731.x}{\emph{MNRAS} {\bfseries 411} (2011) 955} [\href{https://arxiv.org/abs/1003.3878}{{\ttfamily 1003.3878}}].

\end{thebibliography}\endgroup

\end{document}